\documentclass[aps,pra,reprint,amsmath,amssymb,groupedaddress,showpacs]{revtex4-2}
\usepackage{graphicx}                  
\usepackage{float}
\usepackage{dcolumn}                   
\usepackage{bm}                        
\usepackage[bookmarks=false]{hyperref} 
\usepackage[font={small}]{caption}     
\usepackage{threeparttable}            
\usepackage{adjustbox}                 
\usepackage{enumitem}                  
\usepackage{textcomp}                  
\usepackage{xcolor}                    
\hypersetup{
    unicode=false,                     
    pdftoolbar=true,                   
    pdfmenubar=true,                   
    pdffitwindow=true,                 
    pdfstartview={},                   
    pdftitle={Multi-Operator Quantum Uncertainty Relations from New Cauchy-Schwarz Inequalities},          
    pdfauthor={Samuel R. Hedemann},    
    pdfsubject={Quantum Uncertainty Relations},     
    pdfcreator={},                     
    pdfproducer={},                    
    pdfkeywords={Heisenberg Uncertainty Relations,} {Multi-Operator Uncertainty Relations,} {Cauchy-Schwarz Inequalities,} {Multivariance,} {Squeezed States,} {Squeezing},                           
    pdfnewwindow=true,                 
    colorlinks=true,                   
    linkcolor=black,                   
    citecolor=black,                   
    filecolor=black,                   
    urlcolor=black                     
}
\newcommand{\Sec}[1]{\hyperref[sec:#1]{Sec.{\kern 2pt}\ref*{sec:#1}}}
\newcommand{\Section}[1]{\hyperref[sec:#1]{Section~\ref*{sec:#1}}}
\newcommand{\Secs}[2]{\protect\hyperref[sec:#1]{{Secs.{\kern 2pt}\ref*{sec:#1}--\ref*{sec:#2}}}}
\newcommand{\Fig}[2][]{\hyperref[fig:#2]{Fig.{\kern 2pt}\ref*{fig:#2}#1}}
\newcommand{\Figs}[2]{\protect\hyperref[fig:#1]{Figs.{\kern 2pt}\ref*{fig:#1}--\ref*{fig:#2}}}
\newcommand{\Figure}[2][]{\hyperref[fig:#2]{Figure~\ref*{fig:#2}#1}}
\newcommand{\App}[1]{\hyperref[sec:App.#1]{App.{\kern 2pt}\ref*{sec:App.#1}}}
\newcommand{\Appendix}[1]{\hyperref[sec:App.#1]{Appendix~\ref*{sec:App.#1}}}
\newcommand{\EqLab}[1]{\\\noindent\smash{\raisebox{6pt}[0pt][0pt]{\hypertarget{eq:#1}{}}}\vspace{-11pt}}
\newcommand{\Eq}[1]{\protect\hyperlink{eq:#1}{(\ref*{eq.#1})}}
\newcommand{\Eqs}[2]{\protect\hyperlink{eq:#1}{(\ref*{eq.#1}--\ref*{eq.#2})}}
\newenvironment{Equation}[1]{\EqLab{#1}\begin{equation}\label{eq.#1}}{\end{equation}\par\noindent\ignorespacesafterend}
\newcommand{\Table}[2][]{\hyperref[tab:#2]{Table~\ref*{tab:#2}#1}}
\newcommand{\Tables}[2]{\hyperref[tab:#1]{Tables~\ref*{tab:#1}--\ref*{tab:#2}}}
\newcommand{\ThmLab}[1]{\noindent\smash{\raisebox{11pt}[0pt][0pt]{\hypertarget{Thm:#1}{}}}\textbf{Theorem{\kern 3pt}#1:~~}}
\newcommand{\Thm}[1]{\protect\hyperlink{Thm:#1}{Theorem{\kern 3pt}#1}}
\newcommand{\Thms}[2]{\protect\hyperlink{Thm:#1}{Theorems{\kern 3pt}#1--#2}}
\newcommand{\tr}[0]{{\rm tr}}
\newcommand{\shiftmath}[2]{\textnormal{\raisebox{#1}[0pt][0pt]{$#2$}}}
\newcommand{\scalemath}[2]{\textnormal{\scalebox{#1}{$#2$}}}
\newcommand{\hsp}[1]{{\kern #1pt}}
\renewcommand{\geq}[0]{\geqslant}
\renewcommand{\ge}[0]{\geqslant}

\newcommand{\redx}[2]{{#1}{\kern -5.3pt}{{\textnormal{\raisebox{-1.2pt}{\scalebox{1.2}{\textasciicaron}}}}}{\kern -4.2pt}{~}^{(#2)}}
\newcommand{\redX}[2]{{#1}{\kern -6.9pt}{{\textnormal{\raisebox{1.2pt}{\scalebox{1.2}{\textasciicaron}}}}}{\kern -3.1pt}{~}^{(#2)}}
\newcommand{\mbar}[0]{\mathop {m}\limits^{{\kern -3.5pt}~_{\overline{{\kern 7pt}}}}}
\newcommand{\mbarsub}[0]{{\kern 0.0pt}\mathop {m}\limits^{{\kern -0.3pt}{\overline{{\kern 5.5pt}}}}{\kern 0.0pt}}
\newcommand{\nmaxnot}{n_{\,\overline{{\kern -1.8pt}\max^{~^{~^{~}}}\!\!\!\!\!\!\!\!\!\!}}\,}
\makeatletter
\newcommand{\mydotfill}{\leavevmode \cleaders \hb@xt@ .65em{\hss .\hss }\hfill \kern \z@}
\makeatother
\newcommand{\TOConeRaw}[4]{\hyperlink{#1}{\textbf{#3}}&\hyperlink{#1}{\textbf{#4}}\mydotfill\!\! &\hspace{\stretch{1}}\hyperlink{#1}{\textbf{\pageref*{#2}}}}
\newcommand{\TOConeSubRaw}[4]{\hyperlink{#1}{{\kern 7pt}\textbf{#3}}&\hyperlink{#1}{\textbf{#4}}\mydotfill\!\! &\hspace{\stretch{1}}\hyperlink{#1}{\textbf{\pageref*{#2}}}}
\newcommand{\TOCtwoSubRaw}[4]{\hyperlink{#1}{{\kern 7pt}\textbf{#3}}&\hyperlink{#1}{\textbf{#4}}\mydotfill\!\! &\hspace{\stretch{1}}\raisebox{-10.8pt}{\hyperlink{#1}{\textbf{\pageref*{#2}}}}}
\newcommand{\TOCthreeSubRaw}[4]{\hyperlink{#1}{{\kern 7pt}\textbf{#3}}&\hyperlink{#1}{\textbf{#4}}\mydotfill\!\! &\hspace{\stretch{1}}\raisebox{-20.9pt}{\hyperlink{#1}{\textbf{\pageref*{#2}}}}}
\newcommand{\TOConeAppRaw}[4]{\hyperlink{#1}{{\kern 7pt}\textbf{#3}}&\hyperlink{#1}{\textbf{#4}}\mydotfill\!\! &\hspace{\stretch{1}}\hyperlink{#1}{\textbf{\pageref*{#2}}}}
\newcommand{\TOCtwoAppRaw}[4]{\hyperlink{#1}{{\kern 7pt}\textbf{#3}}&\hyperlink{#1}{\textbf{#4}}\mydotfill\!\! &\hspace{\stretch{1}}\raisebox{-10.8pt}{\hyperlink{#1}{\textbf{\pageref*{#2}}}}}
\newcommand{\TOCthreeAppRaw}[4]{\hyperlink{#1}{{\kern 7pt}\textbf{#3}}&\hyperlink{#1}{\textbf{#4}}\mydotfill\!\! &\hspace{\stretch{1}}\raisebox{-20.9pt}{\hyperlink{#1}{\textbf{\pageref*{#2}}}}}
\newcommand{\TOCone}[2]{\TOConeRaw{Sec:#1}{sec:#1}{#1.}{#2}\\}

\newcommand{\TOConeSub}[3]{\TOConeSubRaw{Sec:#1.#2}{sec:#1.#2}{#2.}{#3}\\}
\newcommand{\TOCtwoSub}[3]{\TOCtwoSubRaw{Sec:#1.#2}{sec:#1.#2}{#2.}{#3}\\}

\newcommand{\TOCAppHeader}[2]{\TOConeRaw{Sec:#1}{sec:App.A}{App.}{#2}\\}

\newenvironment{MyTOC}[1]%
{\noindent\ignorespaces %
\begin{table}[H]
\begin{adjustbox}{left,height=#1\height}
\renewcommand{\arraystretch}{1.0}
\begin{tabular}[b]{l@{\kern 2.5pt}p{0.87\linewidth}@{}p{0.05\linewidth}}
}%
{\end{tabular}{\kern -10pt}%
\end{adjustbox}
\end{table}{\kern -10pt}%
\par\noindent\ignorespacesafterend}%
\newcommand{\TOCSecTarget}[2]{
\begin{figure}[H]
\centering%
\vspace{-12pt}
\setlength{\unitlength}{0.01\linewidth}
\begin{picture}(100,0)
\put(1,24){\hypertarget{Sec:#1}{}}
\end{picture}
\end{figure}
\vspace{#2}
}
\newcommand{\TOCAppHeaderTarget}[2]{
\begin{figure}[H]
\centering%
\vspace{-12pt}
\setlength{\unitlength}{0.01\linewidth}
\begin{picture}(100,0)
\put(1,24){\hypertarget{Sec:#1}{}\hypertarget{Sec:App.A}{}}
\end{picture}
\end{figure}
\vspace{#2}
}
\begin{document}
\title{\mbox{Multi-Operator\hsp{-1.6} Quantum\hsp{-1.6} Uncertainty\hsp{-1.6} Relations\hsp{-1.6} from\hsp{-1.6} New\hsp{-1.6} Cauchy-Schwarz\hsp{-1.6} Inequalities}}
\author{Samuel R.\! Hedemann}
\affiliation{Department of Physics, Stevens Institute of Technology, Hoboken, NJ 07030, USA}
\date{v2: February 22, 2026{\kern 2.5pt}\textbar{\kern 2.5pt}v1: May 30, 2025}
\begin{abstract}
We present new generalizations of Cauchy-Schwarz (CS) inequalities to multiple vectors and use them to derive multi-operator quantum uncertainty relations and propose multi-operator squeezing.
\end{abstract}
\maketitle
\section{\label{sec:I}Introduction}
\TOCSecTarget{I}{-39pt}
Recently, there has been increased interest in finding a multi-operator version \cite{BrCM,MiIm,KeWe,Bann,LiQi,QiFL,HoHe,BaPa,SLPQ,Dodo,ChLi,HeHo} of the Heisenberg uncertainty relation \cite{Hei1,Hei2} and its related forms, such as the Robertson and Schr{\"o}dinger uncertainty relations \cite{Kenn,Weyl,Robe,ScU1,ScU2},
\begin{Equation}                      {1}
\begin{array}{*{20}{c}}
{\begin{array}{*{20}{l}}
{{\text{1927\! --\! Heisenberg:}}}&\;\!\!{\delta x\delta {p_x} \approx \hbar,\,[x,{p_x}]\! =\! i\hbar, }\\[2pt]
{{\text{1928\! --\! Kennard,\! Weyl:}}}&\;\!\!{{\sigma _x}{\sigma _{{p_x}}}\hsp{-0.6} \ge \frac{\hbar }{2},}\\[2pt]
{{\text{1929\! --\! Robertson:}}}&\;\!\!{{\sigma _A}{\sigma _B} \ge \hsp{-1.9}\left| {\frac{{\langle [A,B]\rangle }}{{2i}}} \right|\!,}\\[4.0pt]
{{\text{1930\! --\! Schr{\"o}dinger:}}}&\;\!\!{{\sigma _A}{\sigma _B} \ge {\left|{\langle\scalemath{0.76}{ (A\!-\!\langle A\rangle)(B\!-\!\langle B\rangle)}\rangle}\right|},}
\end{array}}\\[25pt]
{{\sigma _A}{\sigma _B} \ge {\sqrt {{{\left| {\frac{{\langle \{ A,B\} \rangle }}{2} - \langle A\rangle \langle B\rangle } \right|}^2}\! + {{\left| {\frac{{\langle [A,B]\rangle }}{{2i}}} \right|}^2}}} ,}
\end{array}\!\!\!\!
\end{Equation}
where $\delta A$ is an approximate uncertainty, $x$ is position, $p_x$ is $x$-direction momentum, $A$ and $B$ are Hermitian operators, $\sigma_A$ is a standard deviation, $[A,B]\equiv AB-BA$, $\{A,B\}\equiv AB+BA$, and $\langle A\rangle$ is an expectation value \cite{Stei}.

However, these new attempts focus on \textit{tightening} inequalities, and some are based on unproven generalizations of the Cauchy-Schwarz (CS) inequality \cite{Cauc}, such as in \cite{QiFL} which claims their three-operator relation is from the ``generalized Schr{\"o}dinger uncertainty relation'' \textit{without proof}, and claims that there are ``no relations like [Cauchy-]Schwartz inequality for three or more objects.''

The main result of our paper is to derive general multiple-vector CS inequalities and apply them to derive quantum uncertainty relations. (Incidentally, this proves the unproven generalization of the Schr{\"o}dinger uncertainty relation used in \cite{QiFL}, but was found before seeing that paper.)  Furthermore, we point out that pursuing tight inequalities is not necessary because \textit{the actual product of uncertainties itself is the tightest bound, and can already always be calculated}. Therefore, the true usefulness of quantum uncertainty relations as inequalities is in their simplicity, which our results achieve. We then suggest new forms of multi-operator squeezing. \textit{Contents}:\\\vspace{-14pt}
\begin{MyTOC}{0.96}
\TOCone{I}{Introduction}
\TOCone{II}{Summary of Results}
    \TOConeSub{II}{A}{New CS Inequalities for Complex Vectors}
    \TOConeSub{II}{B}{New CS Inequalities for Quantum States}
    \TOCtwoSub{II}{C}{\mbox{Multi-Operator Quantum Uncertainty} \mbox{Relations for Hermitian Operators}}
    \TOCtwoSub{II}{D}{\mbox{Multi-Operator Quantum Uncertainty} \mbox{Relations for Any Operators and States}}
    \TOConeSub{II}{E}{Multivariance Uncertainty Relations}
    \TOConeSub{II}{F}{Multi-Operator Squeezing Definition}
\TOCone{III}{Proofs and Derivations}
\TOCone{IV}{Multi-Operator Squeezing}
\TOCone{V}{Conclusions}
\TOCAppHeader{VI}{Adapting These Results to Mixed States}
\end{MyTOC}
\section{\label{sec:II}Summary of Results}
\TOCSecTarget{II}{-39pt}
Here we briefly list our main results with simple examples. Proofs of these results are in \Sec{III} and \Sec{IV}.

In anticipation of the quantum uncertainty relations, we will put the larger side of the CS relations on the left. In all cases, we refer to any set of $M$ complex-valued $n$-dimensional vectors $\mathbf{a}_{1},\ldots,\mathbf{a}_{M}$ with Hermitian inner product $\mathbf{u}\cdot\mathbf{v}\equiv \mathbf{u}^{\dag}\mathbf{v}$ with Hermitian conjugate $\mathbf{u}^{\dag}\equiv\mathbf{u}^{T*}=\mathbf{u}^{*T}$, which simplifies to the Euclidean inner product when the vectors are all real. Note that although several CS inequality generalizations have been discovered \cite{Call,MaDS,Cho,Bann,Yin1,Yin2}, ours are distinct from those.  See the \hyperlink{Sec:App.A}{Appendix} to adapt these results to mixed states.\\\vspace{-20pt}%
\subsection{\label{sec:II.A}Multiple-Vector Cauchy-Schwarz Inequalities for Complex Vectors}
\TOCSecTarget{II.A}{-39pt}
A \textit{balanced multiple-vector CS inequality} involving all vector magnitudes exactly once on the left is
\begin{Equation}                      {2}
\setlength\fboxsep{4pt}   
\setlength\fboxrule{0.5pt}
\fbox{$\prod\limits_{j = 1}^M {|{{\bf{a}}_j}|}  \ge {\left( {\prod\limits_{j = 1}^{M - 1} {\prod\limits_{k = j + 1}^M {|{{\bf{a}}_j} \cdot {{\bf{a}}_k}|} } } \right)^{\frac{1}{{M - 1}}}}$}\;,%
\end{Equation}
valid for $M \ge 2$ complex vectors, see \Sec{III.A.2} for proof.  For example, for $M = 3$ vectors ${\bf{a}},{\bf{b}},{\bf{c}}$, \Eq{2} yields
\begin{Equation}                      {3}
|{\bf{a}}||{\bf{b}}||{\bf{c}}| \ge \sqrt {|{\bf{a}} \cdot {\bf{b}}||{\bf{a}} \cdot {\bf{c}}||{\bf{b}} \cdot {\bf{c}}|} .
\end{Equation}
For $M = 4$ vectors ${\bf{a}},{\bf{b}},{\bf{c}},{\bf{d}}$, \Eq{2} gives
\begin{Equation}                      {4}
|{\bf{a}}||{\bf{b}}||{\bf{c}}||{\bf{d}}| \ge\sqrt[3]{{|{\bf{a}} \cdot {\bf{b}}||{\bf{a}} \cdot {\bf{c}}||{\bf{a}} \cdot {\bf{d}}||{\bf{b}} \cdot {\bf{c}}||{\bf{b}} \cdot {\bf{d}}||{\bf{c}} \cdot {\bf{d}}|}}.
\end{Equation}

A more general \textit{unbalanced multiple-vector CS inequality} for any monomial combination of powers of vector magnitudes of any $K\ge 1$ pairs of vectors on the left is
\begin{Equation}                      {5}
\setlength\fboxsep{4pt}   
\setlength\fboxrule{0.5pt}
\fbox{$\prod\limits_{q = 1}^K {|{{\bf{a}}_{{j_q}}}||{{\bf{a}}_{{k_q}}}|}  \ge \prod\limits_{q = 1}^K {|{{\bf{a}}_{{j_q}}} \cdot {{\bf{a}}_{{k_q}}}|} $}\;,
\end{Equation}
where each $q$th pair ${{\bf{a}}_{{j_q}}},{{\bf{a}}_{{k_q}}}$ can have vectors in common with other pairs; see \Sec{III.A.1} for proof.\hsp{-2}  For example, for $K\in 1,2,3$ with various vector definitions, \Eq{5} yields
\begin{Equation}                      {6}
\begin{array}{rcl}
{|{\bf{a}}{|^2}|{\bf{b}}{|^2}|{\bf{c}}{|^2}}&\!\! \ge &\!\!{|{\bf{a}} \cdot {\bf{b}}||{\bf{a}} \cdot {\bf{c}}||{\bf{b}} \cdot {\bf{c}}|,}\\
{|{\bf{a}}{|^2}|{\bf{b}}||{\bf{c}}|}&\!\! \ge &\!\!{|{\bf{a}} \cdot {\bf{b}}||{\bf{a}} \cdot {\bf{c}}|,}\\
{|{\bf{a}}||{\bf{b}}{|^2}|{\bf{c}}|}&\!\! \ge &\!\!{|{\bf{a}} \cdot {\bf{b}}||{\bf{b}} \cdot {\bf{c}}|,}\\
{|{\bf{a}}||{\bf{b}}||{\bf{c}}{|^2}}&\!\! \ge &\!\!{|{\bf{a}} \cdot {\bf{c}}||{\bf{b}} \cdot {\bf{c}}|,}\\
{|{\bf{a}}||{\bf{b}}| \ge |{\bf{a}} \cdot {\bf{b}}|,\quad |{\bf{a}}||{\bf{c}}|}&\!\! \ge &\!\!{|{\bf{a}} \cdot {\bf{c}}|,\quad |{\bf{b}}||{\bf{c}}| \ge |{\bf{b}} \cdot {\bf{c}}|,}
\end{array}
\end{Equation}
where the top line of \Eq{6} is equivalent to \Eq{3} and the bottom line is equivalent to the original two-vector CS inequality.  Thus \Eq{5} contains \Eq{2} as special cases.

\newpage
More generally,\hsp{-1} for any complex vector\hsp{-1} $\mathbf{a}$\hsp{-1} and\hsp{-1} $M\!\geq\! 2$ $n\times n$ Hermitian matrices $ A_1, \ldots ,A_M$, if \smash{${\mathbf{a}_{{j_1}, \ldots ,{j_q}}} \equiv$} \smash{$[\prod\nolimits_{r\, =\, 1}^q {(\Delta {A_{{j_r}}})} ]\mathbf{a}$}, where $\Delta {A} \!\equiv\! A \!-\! \langle {A}\rangle $ and $\langle {A}\rangle  \!\equiv\! {\mathbf{a}^\dag }{A}\mathbf{a}=\tr(\mathbf{a}\mathbf{a}^{\dag}A)$, then we get \textit{multivariance CS inequalities}
\begin{Equation}                      {7}
\setlength\fboxsep{4pt}   
\setlength\fboxrule{0.5pt}
\fbox{$|{\mathbf{a}_{{p}, \ldots ,{1}}}||{\mathbf{a}_{{{p + 1}}, \ldots {M}}}| \ge |\sigma _{{A_1}, \ldots ,{A_M}}^M|$}\;,
\end{Equation}
where $p \in 0, \ldots ,M$, subscripts in the left-side factors are generally descending and ascending nonoverlapping subsets of $\{0, \ldots ,M+1\}$, we let ${\mathbf{a}_{{0}, \ldots ,{1}}} \equiv \mathbf{a}$ and ${\mathbf{a}_{{{M+1}}, \ldots ,{M}}} \equiv \mathbf{a}$ [see \Eqs{89}{92}], and we define the \textit{multivariance} as
\begin{Equation}                      {8}
\setlength\fboxsep{4pt}   
\setlength\fboxrule{0.5pt}
\fbox{$\sigma _{{A_1}, \ldots ,{A_M}}^M \equiv {\mathop{\rm cov}} ({A_1}, \ldots ,{A_M}) \equiv \left\langle {\prod\limits_{j = 1}^M {(\Delta {A_j})} } \right\rangle $}\;,
\end{Equation}
as proved in \Sec{III.D}. For example, for $M=2$, \Eq{7} gives
\begin{Equation}                      {9}
\begin{array}{*{20}{l}}
{|\mathbf{a}||{\mathbf{a}_{1,2}}|}&\!\!{ \ge |\sigma _{{A_1},{A_2}}^2|,}\\
{|{\mathbf{a}_1}||{\mathbf{a}_2}|}&\!\!{ \ge |\sigma _{{A_1},{A_2}}^2|,}\\
{|\mathbf{a}_{2,1}||{\mathbf{a}}|}&\!\!{ \ge |\sigma _{{A_1},{A_2}}^2|,}
\end{array}
\end{Equation}
and for $M=3$, \Eq{7} yields
\begin{Equation}                      {10}
\begin{array}{*{20}{l}}
{|\mathbf{a}||{\mathbf{a}_{1,2,3}}|}&\!\!{ \ge |\sigma _{{A_1},{A_2},{A_3}}^3|,}\\
{|{\mathbf{a}_1}||{\mathbf{a}_{2,3}}|}&\!\!{ \ge |\sigma _{{A_1},{A_2},{A_3}}^3|,}\\
{|{\mathbf{a}_{2,1}}||{\mathbf{a}_3}|}&\!\!{ \ge |\sigma _{{A_1},{A_2},{A_3}}^3|,}\\
{|\mathbf{a}_{3,2,1}||{\mathbf{a}}|}&\!\!{ \ge |\sigma _{{A_1},{A_2},{A_3}}^3|,}
\end{array}
\end{Equation}
and convex combinations of these can yield others. [The definition in \Eq{8} depends on its arguments' \textit{order} to reveal the most detail. See \Sec{V} for its symmetrized form.]
\\\vspace{-18pt}%
\subsection{\label{sec:II.B}Multiple-Vector Cauchy-Schwarz Inequalities for Quantum States}
\TOCSecTarget{II.B}{-39pt}
Expressing \Eq{2} in terms of pure quantum states (shown here not necessarily normalized to $1$) gives
\begin{Equation}                      {11}
\setlength\fboxsep{4pt}   
\setlength\fboxrule{0.5pt}
\fbox{$\prod\limits_{j = 1}^M {\sqrt {\langle {\psi _j}|{\psi _j}\rangle } }  \ge {\left( {\prod\limits_{j = 1}^{M - 1} {\prod\limits_{k = j + 1}^M {|\langle {\psi _j}|{\psi _k}\rangle |} } } \right)^{\frac{1}{{M - 1}}}}$}\;,
\end{Equation}
for $M \ge 2$, where the left side is $1$ if all states are already normalized.  Similarly, \Eq{5} becomes
\begin{Equation}                      {12}
\setlength\fboxsep{4pt}   
\setlength\fboxrule{0.5pt}
\fbox{$\prod\limits_{q = 1}^K {\sqrt {\langle {\psi _{{j_q}}}|{\psi _{{j_q}}}\rangle } \sqrt {\langle {\psi _{{k_q}}}|{\psi _{{k_q}}}\rangle } }  \ge \prod\limits_{q = 1}^K {|\langle {\psi _{{j_q}}}|{\psi _{{k_q}}}\rangle |} $}\;,
\end{Equation}
for $K \ge 1$, where the left side is $1$ if all states are already normalized.  Both of these results in \Eq{11} and \Eq{12} come from the inner product's definition for quantum states, and are equivalent to \Eq{2} and \Eq{5}. Proofs in \Sec{III.B}.

As proved in \Sec{III.D}, expressing \Eq{7} in terms of pure, not necessarily normalized quantum states gives
\begin{Equation}                      {13}
\setlength\fboxsep{4pt}   
\setlength\fboxrule{0.5pt}
\fbox{$\rule{0pt}{21pt}\rule[-16.0pt]{0pt}{14.0pt}\scalemath{0.89}{\begin{array}{*{20}{l}}
{\sqrt {\langle {\psi _{{p}, \ldots ,{1}}}|{\psi _{{p}, \ldots ,{1}}}\rangle } \sqrt {\langle {\psi _{{{p + 1}}, \ldots {M}}}|{\psi _{{{p + 1}}, \ldots {M}}}\rangle } }&\!\!{ \ge |\sigma _{{A_1}, \ldots ,{A_M}}^M|}\\[4.5pt]
{\sqrt {\sigma _{{A_{{1}}}, \ldots ,{A_{{p}}},{A_{{p}}}, \ldots ,{A_{{1}}}}^{2p}} \sqrt {\sigma _{{A_{{M}}}, \ldots ,{A_{{{p + 1}}}},{A_{{{p + 1}}}}, \ldots ,{A_{{M}}}}^{2(M-p)}} }&\!\!{ \ge |\sigma _{{A_1}, \ldots ,{A_M}}^M|}
\end{array}}$}\;,
\end{Equation}
where $|{\psi _{j_{1}, \ldots ,j_{q}}}\rangle  \equiv [\prod\nolimits_{r = 1}^q {(\Delta {A_{j_{r}}})} ]|\psi \rangle $, $|{\psi _{{0}, \ldots ,{1}}}\rangle  \equiv |\psi \rangle $, $|{\psi _{{M+1}, \ldots ,{M}}}\rangle  \equiv |\psi \rangle $, and $\sigma_{A_{j_1},\ldots,A_{j_q},A_{j_q},\ldots,A_{j_1}}^{0}\!\equiv \langle\psi|\psi\rangle$.
\subsection{\label{sec:II.C}Multi-Operator Quantum Uncertainty Relations for Hermitian Operators and Normalized States}
\begin{figure}[H]
\centering
\includegraphics[width=1.00\linewidth]{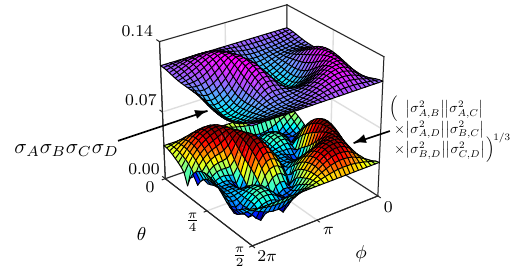}
\vspace{-14pt}
\caption[]{Example of the balanced multi-operator uncertainty relation from \Eq{14} for $M=4$ arbitrary Hermitian operators, for one-qubit mixed states $\rho\equiv \epsilon\Lambda\epsilon^{\dag}$ where $\Lambda\equiv\text{diag}\{\lambda_{1},\lambda_{2}\}$ for $\lambda_1 =c_{\pi/8}^2$, $\!\lambda_2 \,=\,s_{\pi/8}^2$ where $c_x\equiv\cos(x)$ and  $s_x\equiv\sin(x)$, and \smash{$\epsilon\!\equiv\!(\!\phantom{.}_{s_{\theta}e^{i\phi}}^{c_\theta}\!\!\phantom{.}_{\phantom{-}c_\theta}^{-s_{\theta}e^{-i\phi}})$}. Color for illustration only.}
\label{fig:1}
\end{figure}

\TOCSecTarget{II.C}{-39pt}
For a system in normalized state $\rho$, mixed or pure, \Eq{11} leads to a \textit{balanced multi-operator uncertainty relation} for any $M$ Hermitian operators ${A_1}, \ldots ,{A_M}$ as
\begin{Equation}                      {14}
\setlength\fboxsep{4pt}   
\setlength\fboxrule{0.5pt}
\fbox{$\prod\limits_{j = 1}^M {{\sigma _{{A_j}}}}  \ge {\left( {\prod\limits_{j = 1}^{M - 1} {\prod\limits_{k = j + 1}^M {|\sigma _{{A_j},{A_k}}^2|} } } \right)^{\frac{1}{{M - 1}}}}$}\;,
\end{Equation}
for $M \ge 2$, where $\sigma _{A,B}^2$ is the \textit{covariance} between Hermitian operators $A$ and $B$, given by
\begin{Equation}                      {15}
\sigma _{A,B}^2 \!\equiv\! \text{cov}(A,\hsp{-1.0}B) \!\equiv\! \left\langle{\rule{0pt}{9pt}\hsp{-1.0} (A\!-\!\langle A\rangle )(B\!-\!\langle B\rangle )\hsp{-1.0}}\right\rangle \hsp{-3.6}=\hsp{-2.5}\langle AB\rangle  \!-\! \langle A\rangle \langle B\rangle \hsp{-0.5},
\end{Equation}
so that the \textit{variance} is
\begin{Equation}                      {16}
\sigma _A^2 \equiv \sigma _{A,A}^2 \equiv  \text{cov}(A,A) \equiv \left\langle{\rule{0pt}{9pt} (A\!-\!\langle A\rangle )^{2}}\right\rangle =\langle {A^2}\rangle  - {\langle A\rangle ^2},
\end{Equation}
so then \smash{${\sigma _A}\equiv\sqrt{\sigma_A^2}$} is the \textit{standard deviation}, and note that the expectation values are state-dependent as
\begin{Equation}                      {17}
\langle C\rangle  \equiv {\langle C\rangle _\rho } \equiv \tr(\rho C),
\end{Equation}
which, for pure states $|\psi \rangle $ simplifies to $\langle C\rangle  \equiv \langle \psi |C|\psi \rangle $.  \Figure{1} shows an example of \Eq{14} for $M=4$ operators.

As another example, for $M = 3$ Hermitian operators $A,B,C$ and normalized state  $|\psi \rangle $, \Eq{14} yields
\begin{Equation}                      {18}
\scalemath{0.89}{\begin{array}{l}
{\sigma _A}{\sigma _B}{\sigma _C}\! \ge\! \sqrt {\textstyle\,\left|{\sigma _{A,B}^2}\right|\!\left|{\sigma _{A,C}^2}\right|\!\left|{\sigma _{B,C}^2}\right|} \\
{\sigma _A}{\sigma _B}{\sigma _C}\! \ge\! \sqrt {\left|{\rule{0pt}{9pt}\langle AB\rangle  \!-\! \langle A\rangle \langle B\rangle }\right|\!\left|{\rule{0pt}{9pt}\langle AC\rangle  \!-\! \langle A\rangle \langle C\rangle }\right|\!\left|{\rule{0pt}{9pt}\langle BC\rangle  \!-\! \langle B\rangle \langle C\rangle }\right|} ,
\end{array}}
\end{Equation}
where it is helpful to note that the factors on the right can be expressed in several different forms, such as
\begin{Equation}                      {19}
\begin{array}{*{20}{l}}
{|\sigma _{A,B}^2|}&\!\!{ \equiv\left|{\left\langle{\rule{0pt}{9pt} (A-\langle A\rangle )(B-\langle B\rangle )}\right\rangle}\right| = \left|{\rule{0pt}{9pt}\langle AB\rangle  - \langle A\rangle \langle B\rangle }\right|}\\[2pt]
{}&\!\!{ = \sqrt {\left({\rule{0pt}{10pt}\langle AB\rangle  - \langle A\rangle \langle B\rangle }\right)\left({\rule{0pt}{10pt}\langle BA\rangle  - \langle A\rangle \langle B\rangle }\right)} }\\
{}&\!\!{ = \sqrt {{{\left| {\frac{{\langle \{ A,B\} \rangle }}{2} - \langle A\rangle \langle B\rangle } \right|}^2} + {{\left| {\frac{{\langle [A,B]\rangle }}{{2i}}} \right|}^2}}. }
\end{array}
\end{Equation}

\begin{figure}[H]
\centering
\includegraphics[width=1.00\linewidth]{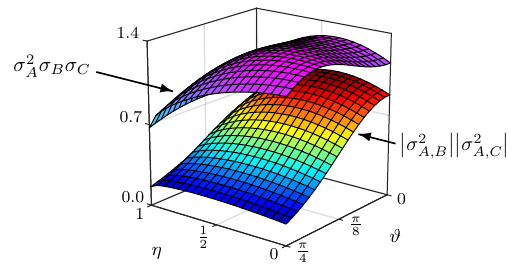}
\vspace{-18pt}
\caption[]{Demonstration of the unbalanced multi-operator uncertainty relation of \Eq{20} for the example in \Eq{21} of three arbitrary Hermitian operators and a family of rank-2 two-qubit states parameterized by spectrum $(\lambda_{1},\lambda_{2})$ and concurrence $C$ \cite{HiWo,Woot} as in \cite{HeXU} as \smash{$(\lambda_{1},\lambda_{2})\equiv(c_{\vartheta}^2,s_{\vartheta}^2)$}, and \smash{$C=\eta c_{\vartheta}^2$} where $\eta\in[0,1]$ and \smash{$\vartheta\in[0,\frac{\pi}{4}]$}. Color for illustration only.}
\label{fig:2}
\end{figure}

Similarly, for a system in normalized state $\rho$, mixed or pure, an \textit{unbalanced multi-operator uncertainty relation} for any $K \ge 1$ pairs of Hermitian operators can be derived from \Eq{12} as
\begin{Equation}                      {20}
\setlength\fboxsep{4pt}   
\setlength\fboxrule{0.5pt}
\fbox{$\prod\limits_{q = 1}^K {{\sigma _{{A_{{j_q}}}}}{\sigma _{{A_{{k_q}}}}}}  \ge \prod\limits_{q = 1}^K {|\sigma _{{A_{{j_q}}},{A_{{k_q}}}}^2|}$}\;,
\end{Equation}
where \Eqs{15}{17} and \Eq{19} all apply as well. For example,
\begin{Equation}                      {21}
\begin{array}{l}
\sigma _A^2{\sigma _B}{\sigma _C} \ge \left|\sigma _{A,B}^2\right|\left|{\sigma _{A,C}^2}\right|\\
\sigma _A^2{\sigma _B}{\sigma _C} \ge \left|{\rule{0pt}{9pt}\langle AB\rangle  - \langle A\rangle \langle B\rangle }\right|\left|{\rule{0pt}{9pt}\langle AC\rangle  - \langle A\rangle \langle C\rangle }\right|,
\end{array}
\end{Equation}
which is again expressible in other ways using \Eq{19}, and a plot of \Eq{21} for three arbitrary operators and a two-qubit state family is shown in \Fig{2}.
\\\vspace{-18pt}

However, the \textit{tightest} result is always the product itself,
\begin{Equation}                      {22}
\setlength\fboxsep{4pt}   
\setlength\fboxrule{0.5pt}
\fbox{$\prod\limits_{j = 1}^M {{\sigma _{{A_j}}}}  = \prod\limits_{j = 1}^M {\sqrt{\langle {A_{j}^{2}}\rangle-\langle A_{j}\rangle^{2}}}$}\;,
\end{Equation}
examples of which are the upper surfaces in \Figs{1}{2}. Proofs for this section are in \Sec{III.C.2}.
\subsection{\label{sec:II.D}Multi-Operator Quantum Uncertainty Relations for Any Operators and Unnormalized States}
\begin{figure}[H]
\centering
\includegraphics[width=1.00\linewidth]{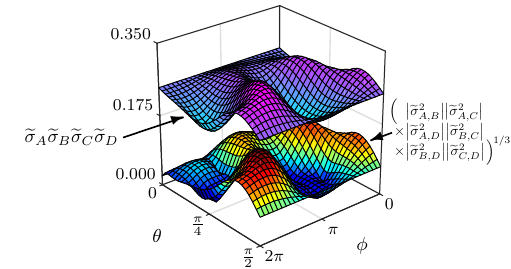}
\vspace{-14pt}
\caption[]{Demonstration of the balanced multi-operator uncertainty relation of \Eq{23} for four arbitrary nonHermitian operators and the same family of one-qubit states used in \Fig{1}. Color for illustration only.}
\label{fig:3}
\end{figure}

\TOCSecTarget{II.D}{-39pt}
For a system in a generally unnormalized state $\rho$, mixed or pure, a \textit{balanced multi-operator uncertainty relation} for any $M$ operators ${A_1}, \ldots ,{A_M}$ (not necessarily Hermitian) can be derived from \Eq{11} as
\begin{Equation}                      {23}
\setlength\fboxsep{4pt}   
\setlength\fboxrule{0.5pt}
\fbox{$\prod\limits_{j = 1}^M {{{\widetilde{\sigma} }_{{A_j}}}}  \ge {\left( {\prod\limits_{j = 1}^{M - 1} {\prod\limits_{k = j + 1}^M {|\widetilde{\sigma} _{{A_j},{A_k}}^2|} } } \right)^{\frac{1}{{M - 1}}}}$}\;,
\\~\\
\end{Equation}
for $M \ge 2$, where we define the \textit{generalized covariance} as
\begin{Equation}                      {24}
\widetilde{\sigma}_{A,B}^2 \equiv \widetilde{\text{cov}}(A,B) \equiv \langle {A^\dag }B\rangle  - [2 - \tr(\rho )]\langle {A^\dag }\rangle \langle B\rangle ,
\end{Equation}
\\
so that the \textit{generalized variance} is
\begin{Equation}                      {25}
\widetilde{\sigma}_A^2 \equiv \widetilde{\sigma}_{A,A}^2 \equiv \widetilde{\text{cov}}(A,A) \equiv \langle {A^\dag }A\rangle  - [2 - \tr(\rho )]\langle {A^\dag }\rangle \langle A\rangle ,
\end{Equation}
and we define the \textit{generalized standard deviation} as
\begin{Equation}                      {26}
{\widetilde{\sigma}_{A}} \equiv \sqrt{\widetilde{\sigma}_{A}^2} \equiv \sqrt{\widetilde{\sigma}_{A,A}^2}\geq 0.
\end{Equation}
We then get alternative special forms of these as
\begin{widetext}~
{\\\vspace{-28.5pt}}%
\begin{Equation}                      {27}
\scalemath{1.00}{\begin{array}{*{20}{l}}
{\left|\widetilde{\sigma} _{A,B}^2\right|}&\!\!{ = \left|{\langle {A^\dag }B\rangle  - [2 - \tr(\rho )]\langle {A^\dag }\rangle \langle B\rangle }\right|= \sqrt {\left({\rule{0pt}{9.5pt}\langle {B^{\dag} }A\rangle  - [2 - \tr(\rho )]\langle A\rangle \langle {B^{\dag} }\rangle }\right)\left({\rule{0pt}{9.5pt}\langle {A^{\dag} }B\rangle  - [2 - \tr(\rho )]\langle {A^{\dag} }\rangle \langle B\rangle }\right)}}\\
{}&\!\!{ = \sqrt {{{\left| {\textstyle\frac{{\langle \,\widetilde{\{} A,B\widetilde{\}}\, \rangle }}{2} - [2 - \tr(\rho )]{\mathop{\rm Re}\nolimits} [\langle {A^\dag }\rangle \langle B\rangle ]} \right|}^2} + {{\left| {\textstyle\frac{{\langle \,\widetilde{[}A,B\widetilde{]}\,\rangle }}{{2i}} - [2 - \tr(\rho )]{\mathop{\rm Im}\nolimits} [\langle {A^\dag }\rangle \langle B\rangle ]} \right|}^2}}, }
\end{array}}
\end{Equation}~
{\\\vspace{-20pt}}
\end{widetext}
which has a much more balanced form in the last line than its analogue of the Schr{\"o}dinger relation from \Eq{1}, and where we define the \textit{pseudocommutator},
\begin{Equation}                      {28}
\widetilde{[}A,B\widetilde{]} \equiv {A^\dag }B - {B^\dag }A,
\end{Equation}
and the \textit{pseudoanticommutator},
\begin{Equation}                      {29}
\widetilde{\{} A,B\widetilde{\}}  \equiv {A^\dag }B + {B^\dag }A,
\end{Equation}
where examples of \Eq{23} can be seen in \Fig{3}. Proofs for this section are in \Sec{III.C.1}.

\begin{figure}[H]
\centering
\includegraphics[width=1.00\linewidth]{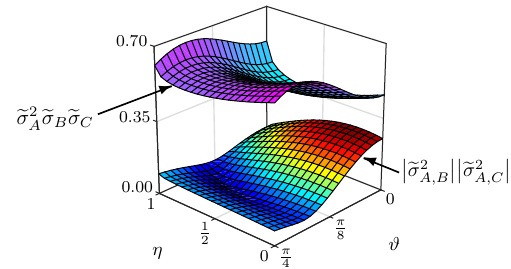}
\vspace{-14pt}
\caption[]{Examples of the unbalanced multi-operator uncertainty relation of \Eq{30} for three arbitrary nonHermitian operators and the same family of two-qubit states used in \Fig{2}. Color for illustration only.}
\label{fig:4}
\end{figure}

Similarly, for a system in unnormalized state $\rho$, mixed or pure, an \textit{unbalanced multi-operator uncertainty relation} for any $K \ge 1$ pairs of not-necessarily Hermitian operators can be derived from \Eq{12} as
\begin{Equation}                      {30}
\setlength\fboxsep{4pt}   
\setlength\fboxrule{0.5pt}
\fbox{$\prod\limits_{q = 1}^K {{{\widetilde{\sigma} }_{{A_{{j_q}}}}}{{\widetilde{\sigma} }_{{A_{{k_q}}}}}}  \ge \prod\limits_{q = 1}^K {|\widetilde{\sigma} _{{A_{{j_q}}},{A_{{k_q}}}}^2|}$}\;,
\end{Equation}
where \Eqs{24}{29} also apply, with examples shown in \Fig{4}.

Finally, similarly to \Eq{22}, the \textit{tightest} relation comes from the uncertainty product itself, as
\begin{Equation}                      {31}
\setlength\fboxsep{4pt}   
\setlength\fboxrule{0.5pt}
\fbox{$\prod\limits_{j = 1}^M {{\widetilde{\sigma}_{{A_j}}}}  = \prod\limits_{j = 1}^M {\sqrt{\langle {A_{j}^\dag }A_{j}\rangle  - [2 - \tr(\rho )]\langle {A_{j}^\dag }\rangle \langle A_{j}\rangle}}$}\;,
\end{Equation}
with examples seen in the upper surfaces in \Figs{3}{4}. Proofs for this section are also in \Sec{III.C}.

It is rare that we might need such quantities as \Eq{23} and \Eq{30}, since most practical work deals with Hermitian observables and normalized states, such as with Bloch vectors \cite{Blch,Stok,VNeu,HiEb,HedD,HCor}.  However, the expectation values of nonHermitian quantities can always be expanded in terms of expectation values of Hermitian operators by using a Hilbert-Schmidt complete Hermitian basis such as the generalized Gell-Mann matrices \cite{Neem,Gell,HedD,HCor} to expand the nonHermitian operators, keeping in mind that they will have complex coefficients in such expansions.

Nevertheless, we include these more general results in case they have some application in the future.
\subsection{\label{sec:II.E}Multivariance Uncertainty Relations}
\begin{figure}[H]
\centering
\includegraphics[width=1.00\linewidth]{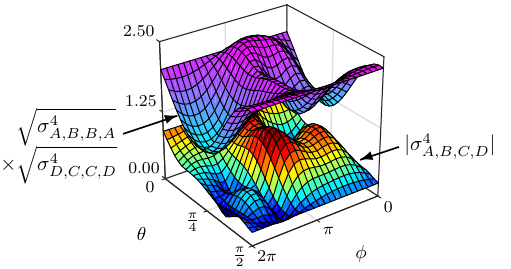}
\vspace{-14pt}
\caption[]{Demonstration of the multivariance relation of \Eq{13} for the same example states and operators as \Fig{1}, for line 3 of \Eq{32}. Color for illustration only.}
\label{fig:5}
\end{figure}

\TOCSecTarget{II.E}{-39pt}
The multivariance uncertainty relations of \Eq{13} are proved in \Sec{III.D}, so here, we simply illustrate them with a nontrivial example for $M=4$ Hermitian operators $A,B,C,D$, which by \Eq{13} produces
\begin{Equation}                      {32}
\begin{array}{*{20}{r}}
{\sqrt{\langle \psi |\psi \rangle} \sqrt {\sigma _{D,C,B,A,A,B,C,D}^8} }&\!\!{ \ge |\sigma _{A,B,C,D}^4|,}\\[5pt]
{\sqrt {\sigma _{A,A}^2} \sqrt {\sigma _{D,C,B,B,C,D}^6} }&\!\!{ \ge |\sigma _{A,B,C,D}^4|,}\\[5pt]
{\sqrt {\sigma _{A,B,B,A}^4} \sqrt {\sigma _{D,C,C,D}^4} }&\!\!{ \ge |\sigma _{A,B,C,D}^4|,}\\[5pt]
{\sqrt {\sigma _{A,B,C,C,B,A}^6} \sqrt {\sigma _{D,D}^2} }&\!\!{ \ge |\sigma _{A,B,C,D}^4|,}\\[5pt]
{\sqrt {\sigma _{A,B,C,D,D,C,B,A}^8} \sqrt{\langle \psi |\psi \rangle} }&\!\!{ \ge |\sigma _{A,B,C,D}^4|,}
\end{array}
\end{Equation}
the third of which is demonstrated in \Fig{5}.
\subsection{\label{sec:II.F}Multi-Operator Squeezing Definition}
\TOCSecTarget{II.F}{-39pt}
In \Sec{IV}, we propose to define $q/M$ \textit{operator squeezing} (or just $q/M$ \textit{squeezing}) for $q \in 1, \ldots ,M - 1$, such that if the lower limit of its generalized variance product (with respect to a particular $M$-operator uncertainty relation) is $\beta$, then using the balanced any-operator uncertainty relation from \Eq{23} for a set of $M$ operators, we say a state is $q/M$ \textit{squeezed} if any particular subset \smash{$\{ A_{j_1}, \ldots ,A_{j_q}\} $} of those operators obeys
\begin{Equation}                      {33}
\setlength\fboxsep{4pt}   
\setlength\fboxrule{0.5pt}
\fbox{$\begin{array}{*{20}{l}}
{{{\widetilde{\sigma} }_{{A_{{j_1}}}}^2}\!\! < {\beta ^{1/M}} \ldots \,\,\text{and}\,\, \ldots\; {{\widetilde{\sigma} }_{{A_{{j_q}}}}^2}\!\! < {\beta ^{1/M}}}\\
{\text{s.t. all uncertainty relations are still obeyed}}
\end{array}$}\;,
\end{Equation}
where, using the square of the right side of \Eq{23},
\begin{Equation}                      {34}
\beta  \equiv {\left( {\prod\limits_{j = 1}^{M - 1} {\prod\limits_{k = j + 1}^M {|\widetilde{\sigma}_{{A_j},{A_k}}^2|} } } \right)^{\!\!\frac{2}{{M - 1}}}},
\end{Equation}
and the ``uncertainty relations still obeyed'' are all balanced any-operator uncertainty relations from \Eq{23} for this $M$ or smaller $M$, where the simplification should be the same (i.e., if using the Robertson version for $M$ operators, use that for all relations of smaller $M$ as well).

For example, for $M=3$ operators $A,B,C$, a state is $2/3$ squeezed if, for constants $a,b,c>1$, and if $\beta\geq 1$,
\begin{Equation}                      {35}
\begin{array}{*{20}{l}}
{{\widetilde{\sigma}_A^2} \!=\! \frac{{{\beta ^{1/3}}}}{a}\!<\!\beta^{1/3}\,\,\text{and}\,\,\widetilde{\sigma}_C^2 \!=\! \frac{{{\beta ^{1/3}}}}{c}\!<\!\beta^{1/3}\;\;\text{s.t.}\,\,\widetilde{\sigma}_B^2 \!=\! ac{\beta ^{1/3}},}\\
{\scalemath{0.75}{\widetilde{\sigma}_A^2\widetilde{\sigma}_B^2 =c\beta^{2/3} \ge |\widetilde{\sigma}_{A,B}^2|,\,\, \widetilde{\sigma}_A^2\widetilde{\sigma}_C^2=\frac{{{\beta ^{2/3}}}}{{ac}} \geq |\widetilde{\sigma}_{A,C}^2|,\,\, \widetilde{\sigma}_B^2\widetilde{\sigma}_C^2 =a\beta^{2/3}\ge |\widetilde{\sigma}_{B,C}^2|} .}
\end{array}
\end{Equation}
See \Sec{IV} for a more detailed look at this example.
\section{\label{sec:III}Proofs and Derivations}
\TOCSecTarget{III}{-39pt}
Here we derive all results summarized in \Sec{II} except multi-operator squeezing, which is treated in \Sec{IV}.
\subsection{\label{sec:III.A}Proof of The Multiple-Vector Cauchy-Schwarz Inequalities for Complex Vectors}
Since the \textit{unbalanced} inequalities of \Eq{5} actually contain the \textit{balanced} inequalities of \Eq{2} as a special case, the unbalanced kind are more general so we start with those.
\subsubsection{\label{sec:III.A.1}Proof of the Generally Unbalanced CS Inequality}
Given any set of $M$ $n$-dimensional complex vectors $\{ {\mathbf{a}_1}, \ldots ,{\mathbf{a}_M}\} $, where ${\mathbf{a}_k} \equiv ({a_{k,1}}, \ldots ,{a_{k,n}})$ for $k \in 1, \ldots ,M$, and the \textit{Hermitian inner product} $\mathbf{u} \cdot \mathbf{v} \equiv {\mathbf{u}^\dag }\mathbf{v}$ with Hermitian conjugate ${\mathbf{u}^\dag } \equiv {\mathbf{u}^{*T}} = {\mathbf{u}^{T*}}$, it is already well-known that the two-vector CS inequality is
\begin{Equation}                      {36}
|{\mathbf{a}_1}||{\mathbf{a}_2}| \ge |{\mathbf{a}_1} \cdot {\mathbf{a}_2}|,
\end{Equation}
which trivially holds if either vector is the zero vector $\mathbf{0}$ as well (we write the product of magnitudes on the left to prepare for the quantum results which are typically written with uncertainty products on the left).

Next, noting that since the overlap magnitude on the right in \Eq{36} involves up to \textit{two} distinct vectors, then there are \smash{$\binom{M}{2} \equiv$} \smash{$ \frac{{M!}}{{2!(M - 2)!}}$} distinct overlap magnitudes. However, since we can pick any subset of those pairs, suppose we make a product of distinct overlap magnitudes between $K \in 1, \ldots ,\binom{M}{2}$ pairs of vectors as
\begin{Equation}                      {37}
\Omega  \equiv \prod\limits_{q = 1}^K {|{\mathbf{a}_{{j_q}}} \cdot {\mathbf{a}_{{k_q}}}|} ,
\end{Equation}
where $\{ {\mathbf{a}_{{j_q}}},{\mathbf{a}_{{k_q}}}\}\equiv {P_q}$ is the $q$th pair of vectors in the chosen set (where curly braces denote a \textit{set} here, not the anticommutator) with vector labels $\{ {j_q},{k_q}\}\equiv L_q $.

For example, for $M=4$ vectors $\{ {\mathbf{a}_1},{\mathbf{a}_2},{\mathbf{a}_3},{\mathbf{a}_4}\} $, there are $\binom{4}{2}=6$ distinct pairs of vectors $\{ \{ {{\bf{a}}_1},{{\bf{a}}_2}\} ,$ $\{ {{\bf{a}}_1},{{\bf{a}}_3}\} ,\{ {{\bf{a}}_1},{{\bf{a}}_4}\} ,\{ {{\bf{a}}_2},{{\bf{a}}_3}\} ,\{ {{\bf{a}}_2},{{\bf{a}}_4}\} ,\{ {{\bf{a}}_3},{{\bf{a}}_4}\} \} $, and if we choose a particular set of $K$ of these as $\{ {P_1},{P_2},{P_3}\}  \equiv \{ \{ {{\bf{a}}_1},{{\bf{a}}_3}\} ,\{ {{\bf{a}}_2},{{\bf{a}}_3}\} ,\{ {{\bf{a}}_2},{{\bf{a}}_4}\} \} $ (letting $q$ label the \textit{new} set's elements), so the chosen label pairs are $\{ {L_1},{L_2},{L_3}\}  \equiv \{ \{ {j_1},{k_1}\} ,\{ {j_2},{k_2}\} ,\{ {j_3},{k_3}\} \} \! \equiv\! \{ \{ 1,3\} ,\{ 2,3\} ,\{ 2,4\} \} $, then \Eq{37} would be $\Omega  \equiv |{{\bf{a}}_1} \cdot {{\bf{a}}_3}||{{\bf{a}}_2} \cdot {{\bf{a}}_3}||{{\bf{a}}_2} \cdot {{\bf{a}}_4}|$ for this subset.

Next, since \Eq{36} also holds in \Eq{37} as
\begin{Equation}                      {38}
|{{\bf{a}}_{{j_q}}}||{{\bf{a}}_{{k_q}}}| \ge |{{\bf{a}}_{{j_q}}} \cdot {{\bf{a}}_{{k_q}}}|;\quad \forall {j_q},{k_q} \in 1, \ldots ,M;\,\,\,q \in 1, \ldots ,K,
\end{Equation}
then putting \Eq{38} into \Eq{37} yields an inequality with $K$ products of magnitudes on the left as
\begin{Equation}                      {39}
\prod\limits_{q = 1}^K {|{{\bf{a}}_{{j_q}}}||{{\bf{a}}_{{k_q}}}|}  \ge \prod\limits_{q = 1}^K {|{{\bf{a}}_{{j_q}}} \cdot {{\bf{a}}_{{k_q}}}|} ,
\end{Equation}
which proves \Eq{5}, while \Eq{6} shows examples of all nontrivial sets of pairs that can be used.  Notice that the product on the left in \Eq{39} can have repeated vector magnitudes from the same vector because multiple different overlap magnitudes on the right could contain the same vector.  In all cases, the powers of each vector magnitude on the left will equal the number of overlap magnitudes that involve that vector on the right, but since not all the powers of all the factors on the left are generally the same, it does not make as much sense to take any radicals except in the special cases where they are all the same.
\subsubsection{\label{sec:III.A.2}Proof of the Balanced CS Inequality}
In the case of the generally unbalanced inequality of \Sec{III.A.1}, if we choose the maximum number of distinct overlap magnitudes of $M$ vectors which is $K=\binom{M}{2}$, then \Eq{39} becomes
\begin{Equation}                      {40}
\prod\limits_{q = 1}^{\binom{M}{2}} {|{{\bf{a}}_{{j_q}}}||{{\bf{a}}_{{k_q}}}|}  \ge \prod\limits_{j = 1}^{M - 1} {\prod\limits_{k = j + 1}^M {|{{\bf{a}}_j} \cdot {{\bf{a}}_k}|} } ,
\end{Equation}
where we cannot just use the same nested product on the left as on the right because the magnitude product would factor and give the wrong number of contributions; this is because on the right, the dot product is generally  a nonfactorizable binary product so $k$ cannot increment without requiring a repeat of the input in the first argument for $j$, and so the same thing must happen on the left of \Eq{40} but a nested product would not achieve that.  However, we can determine the result for the left by looking at the enumerations in the general case as
\begin{Equation}                      {41}
\scalemath{0.80}{\begin{array}{*{20}{l}}
{\prod\limits_{q = 1}^{\binom{M}{2}} {|{{\bf{a}}_{{j_q}}}||{{\bf{a}}_{{k_q}}}|} }&\!\! = &\!\!{{{\left[ {\left( {|{{\bf{a}}_1}||{{\bf{a}}_2}|} \right) \cdots \left( {|{{\bf{a}}_1}||{{\bf{a}}_M}|} \right)} \right]}_1}}\\[-8pt]
{}&\!\!{}&\!\!{ \times {{\left[ {\left( {|{{\bf{a}}_2}||{{\bf{a}}_3}|} \right) \cdots \left( {|{{\bf{a}}_2}||{{\bf{a}}_M}|} \right)} \right]}_2} \cdots {{\left[ {\left( {|{{\bf{a}}_{M - 1}}||{{\bf{a}}_M}|} \right)} \right]}_{M - 1}}}\\[1pt]
{}&\!\! = &\!\!{\!{{\left[ {\,\underline {|{{\bf{a}}_1}{|^{M - 1}}} \left( {\underline{\underline {|{{\bf{a}}_2}|}}  \cdots \underline{\underline {\underline {|{{\bf{a}}_M}|} }} \,} \right)} \right]}_1}}\\[8pt]
{}&\!\!{}&\!\!{{\times\! {\left[ {\,\underline{\underline {|{{\bf{a}}_2}{|^{M - 2}}}} \left( {|{{\bf{a}}_3}| \cdots \underline{\underline {\underline {|{{\bf{a}}_M}|} }} \,} \right)} \right]}_2}\!\!\! \cdots {{\left[ {\left( {|{{\bf{a}}_{M - 1}}|\underline{\underline {\underline {|{{\bf{a}}_M}|} }} \,} \right)} \right]}_{M - 1}}}\\[8pt]
{}&\!\! = &\!\!{\underline {|{{\bf{a}}_1}{|^{(M - 1) + 0}}}\,\, \underline{\underline {|{{\bf{a}}_2}{|^{(M - 2) + 1}}}}  \cdots \underline{\underline {\underline {|{{\bf{a}}_M}{|^{(M - M) + M - 1}}} }} }\\[8pt]
{}&\!\! = &\!\!{|{{\bf{a}}_1}{|^{M - 1}}|{{\bf{a}}_2}{|^{M - 1}} \cdots |{{\bf{a}}_M}{|^{M - 1}}=(|\mathbf{a}_1||\mathbf{a}_2|\cdots|\mathbf{a}_M|)^{M-1}}\\[2pt]
{}&\!\! = &\!\!{{{\left( {\,\smash{\prod\limits_{j = 1}^M {|{{\bf{a}}_j}|}}\rule{0pt}{15pt} } \right)}^{\!\! M - 1}},}
\end{array}}
\end{Equation}
where subscripts on square brackets count groups of magnitude pairs that share the same index in the left factor.  The pattern can be seen by looking at the second equality of \Eq{41} and counting all occurrences of each magnitude, indicated by underline types. The exponent grouping in the third equality shows in the parenthetical exponent on the left the number of common factors appearing on the left in a given square bracket term, while the nonparenthetical exponent on the right shows shows the number of that same factor that are distributed among the other square-bracket factors.  Thus, putting \Eq{41} into \Eq{40} gives
\begin{Equation}                      {42}
{\left( {\,\smash{\prod\limits_{j = 1}^M {|{{\bf{a}}_j}|}}\rule{0pt}{17pt} } \right)^{M - 1}} \ge \prod\limits_{j = 1}^{\! M - 1} {\prod\limits_{k = j + 1}^M {|{{\bf{a}}_j} \cdot {{\bf{a}}_k}|} } ,
\end{Equation}
and then the $(M-1)$th root of both sides of \Eq{42} yields
\begin{Equation}                      {43}
\prod\limits_{j = 1}^M {|{{\bf{a}}_j}|}  \ge {\left( {\prod\limits_{j = 1}^{M - 1} {\prod\limits_{k = j + 1}^M {|{{\bf{a}}_j} \cdot {{\bf{a}}_k}|} } } \right)^{\frac{1}{{M - 1}}}},
\end{Equation}
which proves \Eq{2}, examples of which can be seen in \Eq{3} and \Eq{4}, valid for $M \ge 2$.
\subsection{\label{sec:III.B}Derivation of Multiple-Vector CS Inequalities for Unnormalized Kets}
We show these results separately in \Sec{II.B} because their simplicity might make them easy to get wrong, and for completeness.  First replace all vectors in \Sec{III.A} with $n$-dimensional kets of the same index as
\begin{Equation}                      {44}
{{\bf{a}}_j} \to |{\psi _j}\rangle ,
\end{Equation}
where we will not assume normalized kets here for the sake of generality. Then by the definition of the magnitude of a ket in terms of its self overlap, we have
\begin{Equation}                      {45}
|{{\bf{a}}_j}| = \sqrt {{{\bf{a}}_j} \cdot {{\bf{a}}_j}}  \to \sqrt {\langle {\psi _j}|{\psi _j}\rangle } ,
\end{Equation}
while the overlap between different kets is
\begin{Equation}                      {46}
{{\bf{a}}_j} \cdot {{\bf{a}}_k} \to \langle {\psi _j}|{\psi _k}\rangle ,
\end{Equation}
so putting \Eq{45} and \Eq{46} into \Eq{43} gives
\begin{Equation}                      {47}
\begin{array}{*{20}{c}}
{\prod\limits_{j = 1}^M {|{{\bf{a}}_j}|} }&\!\! \ge &\!\!{{{\left( {\prod\limits_{j = 1}^{M - 1} {\prod\limits_{k = j + 1}^M {|{{\bf{a}}_j} \cdot {{\bf{a}}_k}|} } } \right)}^{\frac{1}{{M - 1}}}}}\\
{\prod\limits_{j = 1}^M {\sqrt {\langle {\psi _j}|{\psi _j}\rangle } } }&\!\! \ge &\!\!{{{\left( {\prod\limits_{j = 1}^{M - 1} {\prod\limits_{k = j + 1}^M {|\langle {\psi _j}|{\psi _k}\rangle |} } } \right)}^{\frac{1}{{M - 1}}}},}
\end{array}
\end{Equation}
which proves \Eq{11}, if all $|{\psi _j}\rangle $ are already normalized, then $\langle {\psi _j}|{\psi _j}\rangle  = 1$, so the whole left side would be $1$ in that case.  Then, using \Eq{45} and \Eq{46} in \Eq{39} gives
\begin{Equation}                      {48}
\begin{array}{*{20}{c}}
{\prod\limits_{q = 1}^K {|{{\bf{a}}_{{j_q}}}||{{\bf{a}}_{{k_q}}}|} }&\!\! \ge &\!\!{\prod\limits_{q = 1}^K {|{{\bf{a}}_{{j_q}}} \cdot {{\bf{a}}_{{k_q}}}|} }\\
{\prod\limits_{q = 1}^K {\sqrt {\langle {\psi _{{j_q}}}|{\psi _{{j_q}}}\rangle } \sqrt {\langle {\psi _{{k_q}}}|{\psi _{{k_q}}}\rangle } } }&\!\! \ge &\!\!{\prod\limits_{q = 1}^K {|\langle {\psi _{{j_q}}}|{\psi _{{k_q}}}\rangle |} ,}
\end{array}
\end{Equation}
which proves \Eq{12}, where again, the left side would be $1$ if all the kets are already normalized.
\subsection{\label{sec:III.C}Proof of Multi-Operator Quantum Uncertainty Relations}
Here, we start by assuming neither Hermiticity of operators nor normalization of kets, which will let us prove the general results in \Sec{II.D}, and then we show that the results in \Sec{II.C} are a special case of that.
\subsubsection{\label{sec:III.C.1}Proof of Multi-Operator Quantum Uncertainty Relations for NonHermitian Operators and Unnormalized States}
First, we define not-necessarily-normalized states
\begin{Equation}                      {49}
|{\psi _k}\rangle  \equiv ({A_k} - \langle {A_k}\rangle )|\psi \rangle ,
\end{Equation}
where $\langle {A_k}\rangle  \equiv \langle \psi |{A_k}|\psi \rangle $ which can be generalized for mixed states $\rho$ as $\langle {A_k}\rangle  \equiv \tr(\rho {A_k})$, and operator $A_k$ is not necessarily Hermitian.  Then, overlaps of $\{|{\psi _k}\rangle\}$ are
\begin{Equation}                      {50}
\scalemath{0.90}{\begin{array}{*{20}{l}}
{\langle {\psi _j}|{\psi _k}\rangle }&\!\!{ = \langle \psi |(A_j^\dag  - {{\langle {A_j}\rangle }^*})({A_k} - \langle {A_k}\rangle )|\psi \rangle }\\[1pt]
{}&\!\!{ = \langle \psi |(A_j^\dag {A_k} - {{\langle {A_j}\rangle }^*}{A_k} - {A_j}^\dag \langle {A_k}\rangle  + {{\langle {A_j}\rangle }^*}\langle {A_k}\rangle )|\psi \rangle }\\[1pt]
{}&\!\!{ = \langle A_j^\dag {A_k}\rangle  \!-\! {{\langle {A_j}\rangle }^*}\langle {A_k}\rangle  \!-\! \langle A_j^\dag \rangle \langle {A_k}\rangle  \!+\! {{\langle {A_j}\rangle }^*}\langle {A_k}\rangle \langle \psi |\psi \rangle. }
\end{array}}\!
\end{Equation}
Then, using the fact that ${\langle {A}\rangle ^*} = \langle A^\dag \rangle $, \Eq{50} becomes
\begin{Equation}                      {51}
\scalemath{0.94}{\begin{array}{*{20}{l}}
{\langle {\psi _j}|{\psi _k}\rangle }&\!\!{ = \langle A_j^\dag {A_k}\rangle  \!-\! \langle A_j^\dag \rangle \langle {A_k}\rangle  \!-\! \langle A_j^\dag \rangle \langle {A_k}\rangle  \!+\! \langle A_j^\dag \rangle \langle {A_k}\rangle \langle \psi |\psi \rangle }\\[1pt]
{}&\!\!{ = \langle A_j^\dag {A_k}\rangle  - 2\langle A_j^\dag \rangle \langle {A_k}\rangle  + \langle A_j^\dag \rangle \langle {A_k}\rangle \tr(\rho )}\\[1pt]
{}&\!\!{ = \langle A_j^\dag {A_k}\rangle  - [2 - \tr(\rho )]\langle A_j^\dag \rangle \langle {A_k}\rangle, }
\end{array}}
\end{Equation}
where we used the standard generalization for kets to density operators \smash{$\rho  \equiv \sum\nolimits_j {{p_j}|{\psi _{\{j\}}}\rangle \langle {\psi _{\{j\}}}|} ;\,\,{p_j} > 0$}, again not necessarily normalized, where we use \smash{$|{\psi _{\{j\}}}\rangle$} to distinguish it from \smash{$|{\psi _j}\rangle$} as in \Eq{49}, and for rank-$1$ states like $|\psi\rangle$, $\rho  = |\psi \rangle \langle \psi |$.  When indices match in \Eq{51}, it becomes
\begin{Equation}                      {52}
\langle {\psi _j}|{\psi _j}\rangle = \langle A_j^\dag {A_j}\rangle  - [2 - \tr(\rho )]\langle A_j^\dag \rangle \langle {A_j}\rangle .
\end{Equation}

Then, we use \Eq{51} to define a \textit{generalized covariance} as
\begin{Equation}                      {53}
\widetilde{\sigma}_{A,B}^2 \equiv \widetilde{\text{cov}}(A,B) \equiv \langle {A^\dag }B\rangle  - [2 - \tr(\rho )]\langle {A^\dag }\rangle \langle B\rangle ,
\end{Equation}
which is the form shown in \Eq{24}, and from \Eq{52} and \Eq{53} we can define a \textit{generalized variance} as
\begin{Equation}                      {54}
\widetilde{\sigma}_A^2 \equiv \widetilde{\sigma}_{A,A}^2 \equiv \widetilde{\text{cov}}(A,A) \equiv \langle {A^\dag }A\rangle  - [2 - \tr(\rho )]\langle {A^\dag }\rangle \langle A\rangle ,
\end{Equation}
as in \Eq{25}. Thus, by \Eq{53} and \Eq{54}, we see that \Eq{51} and \Eq{52} can be written as
\begin{Equation}                      {55}
\!\! \begin{array}{*{20}{l}}
{\langle {\psi _j}|{\psi _k}\rangle }&\!\!{ = \langle A_j^\dag {A_k}\rangle }&\!\!\!{ - [2 - \tr(\rho )]\langle A_j^\dag \rangle \langle {A_k}\rangle }&\!\!{ = \widetilde{\sigma}_{{A_j},{A_k}}^2}\\[1pt]
{\langle {\psi _j}|{\psi _j}\rangle }&\!\!{ = \langle A_j^\dag {A_j}\rangle }&\!\!\!{ - [2 - \tr(\rho )]\langle A_j^\dag \rangle \langle {A_j}\rangle }&\!\!\!{ = \widetilde{\sigma}_{{A_j},{A_j}}^2 \equiv \widetilde{\sigma}_{{A_j}}^2.}
\end{array}
\end{Equation}
Then, putting the results of \Eq{55} into \Eq{47} yields
\begin{Equation}                      {56}
\begin{array}{*{20}{c}}
{\prod\limits_{j = 1}^M {\sqrt {\langle {\psi _j}|{\psi _j}\rangle } } }&\!\! \ge &\!\!{{{\left( {\prod\limits_{j = 1}^{M - 1} {\prod\limits_{k = j + 1}^M {|\langle {\psi _j}|{\psi _k}\rangle |} } } \right)}^{\frac{1}{{M - 1}}}}}\\
{\prod\limits_{j = 1}^M {\sqrt {\widetilde{\sigma}_{{A_j}}^2} } }&\!\! \ge &\!\!{{{\left( {\prod\limits_{j = 1}^{M - 1} {\prod\limits_{k = j + 1}^M {|\widetilde{\sigma}_{{A_j},{A_k}}^2|} } } \right)}^{\frac{1}{{M - 1}}}}}\\
{\prod\limits_{j = 1}^M {{\widetilde{\sigma}_{{A_j}}}} }&\!\! \ge &\!\!{{{\left( {\prod\limits_{j = 1}^{M - 1} {\prod\limits_{k = j + 1}^M {|\widetilde{\sigma}_{{A_j},{A_k}}^2|} } } \right)}^{\frac{1}{{M - 1}}}},}
\end{array}
\end{Equation}
which proves \Eq{23}, and putting \Eq{55} into \Eq{48} gives
\begin{Equation}                      {57}
\begin{array}{*{20}{c}}
{\prod\limits_{q = 1}^K {\sqrt {\langle {\psi _{{j_q}}}|{\psi _{{j_q}}}\rangle } \sqrt {\langle {\psi _{{k_q}}}|{\psi _{{k_q}}}\rangle }} }&\!\! \ge &\!\!{\prod\limits_{q = 1}^K {|\langle {\psi _{{j_q}}}|{\psi _{{k_q}}}\rangle |} }\\
{\prod\limits_{q = 1}^K {\sqrt {\widetilde{\sigma}_{{A_{{j_q}}}}^2} \sqrt {\widetilde{\sigma}_{{A_{{k_q}}}}^2} } }&\!\! \ge &\!\!{\prod\limits_{q = 1}^K {|\widetilde{\sigma}_{{A_{{j_q}}},{A_{{k_q}}}}^2|} }\\
{\prod\limits_{q = 1}^K {{\widetilde{\sigma}_{{A_{{j_q}}}}}{\widetilde{\sigma}_{{A_{{k_q}}}}}} }&\!\! \ge &\!\!{\prod\limits_{q = 1}^K {|\widetilde{\sigma} _{{A_{{j_q}}},{A_{{k_q}}}}^2|}, }
\end{array}
\end{Equation}
which proves \Eq{30}, where we used \smash{${\widetilde{\sigma}_{A}} \equiv \sqrt{\widetilde{\sigma}_{A}^2}$} from \Eq{26}.

To see  how we might get alternative forms of the generalized covariance and variance, first note that overlaps are generally \textit{complex},
\begin{Equation}                      {58}
\begin{array}{*{20}{l}}
{\langle {\psi _j}|{\psi _k}\rangle }&\!\!{ = {\mathop{\rm Re}\nolimits} [\langle {\psi _j}|{\psi _k}\rangle ] + i{\mathop{\rm Im}\nolimits} [\langle {\psi _j}|{\psi _k}\rangle ]}\\
{}&\!\!{ = \frac{{\langle {\psi _j}|{\psi _k}\rangle  + {{\langle {\psi _j}|{\psi _k}\rangle }^*}}}{2} + i\frac{{\langle {\psi _j}|{\psi _k}\rangle  - {{\langle {\psi _j}|{\psi _k}\rangle }^*}}}{{2i}},}
\end{array}
\end{Equation}
Therefore, their \textit{moduli} have the form
\begin{Equation}                      {59}
{\textstyle |\langle {\psi _j}|{\psi _k}\rangle | = \sqrt {{{\left( {\frac{{\langle {\psi _j}|{\psi _k}\rangle  + {{\langle {\psi _j}|{\psi _k}\rangle }^*}}}{2}} \right)}^{\! 2}} \!\!+\! {{\left( {\frac{{\langle {\psi _j}|{\psi _k}\rangle  - {{\langle {\psi _j}|{\psi _k}\rangle }^*}}}{{2i}}} \right)}^{\! 2}}}}.
\end{Equation}
So using \Eq{55} in the first term in the radicand of \Eq{59} as
\begin{Equation}                      {60}
\begin{array}{*{20}{l}}
{\frac{{\langle {\psi _j}|{\psi _k}\rangle  + {{\langle {\psi _j}|{\psi _k}\rangle }^*}}}{2}}&\!\!{ = \frac{{\langle {\psi _j}|{\psi _k}\rangle  + \langle {\psi _k}|{\psi _j}\rangle }}{2}}\\
{}&\!\!{ = \frac{{\langle A_j^\dag {A_k}\rangle  - [2 - \tr(\rho )]\langle A_j^\dag \rangle \langle {A_k}\rangle }}{2}}\\
{}&\!\!{\phantom{=}+\frac{{\langle A_k^\dag {A_j}\rangle  - [2 - \tr(\rho )]\langle A_k^\dag \rangle \langle {A_j}\rangle }}{2}}\\
{}&\!\!{ = \langle \frac{{A_j^\dag {A_k} + A_k^\dag {A_j}}}{2}\rangle }\\
{}&\!\!{\phantom{=}- [2 - \tr(\rho )]\left( {\!\frac{{\langle A_j^\dag \rangle \langle {A_k}\rangle  + \langle A_k^\dag \rangle \langle {A_j}\rangle }}{2}\!} \right)\!,}
\end{array}
\end{Equation}
this prompts us to define a \textit{pseudoanticommutator} as
\begin{Equation}                      {61}
\widetilde{\{}A,B\widetilde{\}} \equiv {A^\dag }B + {B^\dag }A,
\end{Equation}
and if we note that
\begin{Equation}                      {62}
\begin{array}{*{20}{l}}
{\frac{{\langle A_j^\dag \rangle \langle {A_k}\rangle  + \langle A_k^\dag \rangle \langle {A_j}\rangle }}{2}}&\!\!{ = \frac{{{{\langle {A_j}\rangle }^*}\langle {A_k}\rangle  + {{\langle {A_k}\rangle }^*}\langle {A_j}\rangle }}{2}}\\
{}&\!\!{ = {\mathop{\rm Re}\nolimits} [{{\langle {A_j}\rangle }^*}\langle {A_k}\rangle ] = {\mathop{\rm Re}\nolimits} [\langle A_j^\dag \rangle \langle {A_k}\rangle ],}
\end{array}
\end{Equation}
then putting \Eq{61} and \Eq{62} into \Eq{60} gives
\begin{Equation}                      {63}
{\textstyle \frac{{\langle {\psi _j}|{\psi _k}\rangle  + {{\langle {\psi _j}|{\psi _k}\rangle }^*}}}{2} = \langle \frac{{\widetilde{\{} {A_j},{A_k}\widetilde{\}} }}{2}\rangle  - [2 - \tr(\rho )]{\mathop{\rm Re}\nolimits} [\langle A_j^\dag \rangle \langle {A_k}\rangle ]}.
\end{Equation}

Similarly, using \Eq{55} in the second term in \Eq{59},
\begin{Equation}                      {64}
\begin{array}{*{20}{l}}
{\frac{{\langle {\psi _j}|{\psi _k}\rangle  - {{\langle {\psi _j}|{\psi _k}\rangle }^*}}}{{2i}}}&\!\!{ = \frac{{\langle {\psi _j}|{\psi _k}\rangle  - \langle {\psi _k}|{\psi _j}\rangle }}{{2i}}}\\
{}&\!\!{ = \frac{{\langle A_j^\dag {A_k}\rangle  - [2 - \tr(\rho )]\langle A_j^\dag \rangle \langle {A_k}\rangle }}{{2i}}}\\
{}&\!\!{\phantom{=}- \frac{{\langle A_k^\dag {A_j}\rangle  - [2 - \tr(\rho )]\langle A_k^\dag \rangle \langle {A_j}\rangle }}{{2i}}}\\
{}&\!\!{ = \langle \frac{{A_j^\dag {A_k} - A_k^\dag {A_j}}}{{2i}}\rangle }\\
{}&\!\!{\phantom{=}- [2 - \tr(\rho )]\left( {\!\frac{{\langle A_j^\dag \rangle \langle {A_k}\rangle  - \langle A_k^\dag \rangle \langle {A_j}\rangle }}{{2i}}\!} \right)\!,}
\end{array}
\end{Equation}
which prompts us to define a \textit{pseudocommutator} as
\begin{Equation}                      {65}
\widetilde{[} A,B\widetilde{]}  \equiv {A^\dag }B - {B^\dag }A,
\end{Equation}
and if we note that
\begin{Equation}                      {66}
\begin{array}{*{20}{l}}
{\frac{{\langle A_j^\dag \rangle \langle {A_k}\rangle  - \langle A_k^\dag \rangle \langle {A_j}\rangle }}{{2i}}}&\!\!{ = \frac{{{{\langle {A_j}\rangle }^*}\langle {A_k}\rangle  - {{\langle {A_k}\rangle }^*}\langle {A_j}\rangle }}{{2i}}}\\
{}&\!\!{ = {\mathop{\rm Im}\nolimits} [{{\langle {A_j}\rangle }^*}\langle {A_k}\rangle ] = {\mathop{\rm Im}\nolimits} [\langle A_j^\dag \rangle \langle {A_k}\rangle ],}
\end{array}
\end{Equation}
then putting \Eq{65} and \Eq{66} into \Eq{64} gives
\begin{Equation}                      {67}
{\textstyle \frac{{\langle {\psi _j}|{\psi _k}\rangle  - {{\langle {\psi _j}|{\psi _k}\rangle }^*}}}{{2i}} = \langle \frac{{\widetilde{[}{A_j},{A_k}\widetilde{]}}}{{2i}}\rangle  - [2 - \tr(\rho )]{\mathop{\rm Im}\nolimits} [\langle A_j^\dag \rangle \langle {A_k}\rangle ].}
\end{Equation}
Then, noting that due to \Eq{55},
\begin{Equation}                      {68}
|\langle {\psi _j}|{\psi _k}\rangle | = |\widetilde{\sigma}_{{A_j},{A_k}}^2|,
\end{Equation}
putting \Eq{63}, \Eq{67}, and \Eq{68} into \Eq{59} yields
\begin{Equation}                      {69}
{\textstyle |\widetilde{\sigma}_{{A_j},{A_k}}^2| =\! \sqrt{\!\!\begin{array}{l}
{\left( {\frac{{\langle \,\widetilde{\{} {A_j},{A_k}\widetilde{\}}\, \rangle }}{2} - [2 - \tr(\rho )]{\mathop{\rm Re}\nolimits} [\langle A_j^\dag \rangle \langle {A_k}\rangle ]} \right)^2}\\
 + {\left( {\frac{{\langle \,\widetilde{[}{A_j},{A_k}\widetilde{]}\,\rangle }}{{2i}} - [2 - \tr(\rho )]{\mathop{\rm Im}\nolimits} [\langle A_j^\dag \rangle \langle {A_k}\rangle ]} \right)^2}
\end{array}}}, 
\end{Equation}
which proves the third line of \Eq{27} when we set ${A_j} \to A$ and ${A_k} \to B$. Note that \Eq{27} upgrades the squares to square magnitudes to guard against rounding errors in the imaginary parts, but since those quantities are always real due to \Eq{58}, ordinary squares are sufficient.  The first equality in line $1$ of \Eq{27} comes from plugging in \Eq{24}, and the second equality of line $1$ of \Eq{27} just expands the first one by using the modulus rule for complex numbers $|c| = \sqrt{{c^*}c} $ and the facts that ${\langle A\rangle ^*} = \langle {A^\dag }\rangle $ and ${\langle {A^\dag }B\rangle ^*} = \langle {B^\dag }A\rangle $. Thus we have proven everything in \Sec{II.D}.
\subsubsection{\label{sec:III.C.2}Proof of Multi-Operator Quantum Uncertainty Relations for Hermitian Operators and Normalized States}
Now, to prove the results in \Sec{II.C}, we can start with the above results and apply special cases.  First, restricting ourselves to \textit{Hermitian} operators $A^\dag =A$ and \textit{normalized} states $\langle \psi |\psi \rangle  = 1$ and $\tr(\rho ) = 1$, the generalized covariance and generalized variance of \Eq{53} and \Eq{54} simplify to
\begin{Equation}                      {70}
\widetilde{\sigma}_{A,B}^2 \equiv \widetilde{{\mathop{\rm cov}} }(A,B) = \langle AB\rangle  - \langle A\rangle \langle B\rangle  = {\mathop{\rm cov}} (A,B) = \sigma _{A,B}^2 ,
\end{Equation}
which is the familiar \textit{covariance} from \Eq{15}, and
\begin{Equation}                      {71}
\widetilde{\sigma}_A^2 \equiv \widetilde{\sigma}_{A,A}^2 \equiv \widetilde {{\mathop{\rm cov}} }(A,A) = \langle {A^2}\rangle  - {\langle A\rangle ^2} = {\mathop{\rm cov}} (A,A) = \sigma _A^2 ,
\end{Equation}
which is the familiar variance from \Eq{16}.  Then, using \Eq{70} and \Eq{71}, we see that \Eq{56} simplifies to
\begin{Equation}                      {72}
\prod\limits_{j = 1}^M {{\sigma _{{A_j}}}}  \ge {\left( {\prod\limits_{j = 1}^{M - 1} {\prod\limits_{k = j + 1}^M {|\sigma _{{A_j},{A_k}}^2|} } } \right)^{\frac{1}{{M - 1}}}},
\end{Equation}
proving \Eq{14}.  Similarly, using \Eq{70} and \Eq{71} in \Eq{57} gives
\begin{Equation}                      {73}
\prod\limits_{q = 1}^K {{\sigma _{{A_{{j_q}}}}}{\sigma _{{A_{{k_q}}}}}}  \ge \prod\limits_{q = 1}^K {|\sigma _{{A_{{j_q}}},{A_{{k_q}}}}^2|} ,
\end{Equation}
which proves \Eq{20}.  Note that in \Eq{72} and \Eq{73}, there is no need for absolute value bars around the standard deviations, because the square of them, which is the variance, will always be nonnegative due to the fact that it can be written as the mean of a square of a real number due to the Hermiticity of the operators in this case and the fact that it also takes the form $\sigma _A^2 = \langle {(A - \langle A\rangle )^2}\rangle $.

Similarly, for Hermitian operators, our pseudoanticommutator becomes the regular anticommutator as
\begin{Equation}                      {74}
\widetilde{\{} A,B\widetilde{\}}  \equiv {A^\dag }B + {B^\dag }A = AB + BA = \{ A,B\} ,
\end{Equation}
and our pseudocommutator becomes the commutator as
\begin{Equation}                      {75}
\widetilde{[}A,B\widetilde{]} \equiv {A^\dag }B - {B^\dag }A = AB - BA = [A,B].
\end{Equation}
Then, noting that expectation values of Hermitian operators are always \textit{real}, we have
\begin{Equation}                      {76}
\begin{array}{*{20}{l}}
{{\mathop{\rm Re}\nolimits} [\langle A_j^\dag \rangle \langle {A_k}\rangle ]}&\!\!{ = {\mathop{\rm Re}\nolimits} [\langle {A_j}\rangle \langle {A_k}\rangle ]}&\!\!{ = \langle {A_j}\rangle \langle {A_k}\rangle ,}\\
{{\mathop{\rm Im}\nolimits} [\langle A_j^\dag \rangle \langle {A_k}\rangle ]}&\!\!{ = {\mathop{\rm Im}\nolimits} [\langle {A_j}\rangle \langle {A_k}\rangle ]}&\!\!{ = 0,}
\end{array}
\end{Equation}
and again for normalized states,
\begin{Equation}                      {77}
\tr(\rho ) = 1,
\end{Equation}
so putting \Eqs{74}{77} into the right of \Eq{69} and using \Eq{70} on the left of \Eq{69}, we obtain
\begin{Equation}                      {78}
{\textstyle |\sigma _{{A_j},{A_k}}^2| \!=\! \sqrt {{{\left( {\frac{{\langle \{ {A_j},{A_k}\} \rangle }}{2} - \langle {A_j}\rangle \langle {A_k}\rangle } \right)}^2} \!\!+\! {{\left( {\frac{{\langle [{A_j},{A_k}]\rangle }}{{2i}}} \right)}^2}}} ,
\end{Equation}
which proves line $3$ of \Eq{19}, and where again we do not strictly need to use square magnitudes in each term, but we do it to protect against rounding error in imaginary parts of those quantities (which should always be $0$), but also because some people prefer to omit the $i$ in the denominator, which would make the remaining term strictly imaginary, and then the square magnitude would be necessary to make the whole quantity correct.  Lines $1$ and $2$ of \Eq{19} follow from \Eq{27} by specifying that its operators are Hermitian and the state is normalized.
\subsection{\label{sec:III.D}Proofs of Multivariance Uncertainty Relations}
~\\\vspace{-30pt}\\%
In Schr{\"o}dinger's description, the expectation value of a product of two Hermitian operators' \textit{deviations}, defined as $\Delta A\equiv A-\langle A\rangle$, gives the covariance as an \textit{overlap} as
\begin{Equation}                      {79}
\begin{array}{*{20}{l}}
{\langle {\psi _j}|{\psi _k}\rangle }&\!\!{ = \langle \psi |(\Delta{A_j})(\Delta{A_k})|\psi \rangle }\\
{}&\!\!{ = \langle {A_j}{A_k}\rangle  - \langle {A_j}\rangle \langle {A_k}\rangle  \equiv {\mathop{\rm cov}} ({A_j},{A_k}) \equiv \sigma _{{A_j},{A_k}}^{2}.}
\end{array}
\end{Equation}
But for three operators or more, the mean of a product of those operators' deviations is not uniquely describable as an \textit{overlap} of just two states.  For example,
\begin{Equation}                      {80}
\begin{array}{*{20}{l}}
{\langle \psi |(\Delta{A_j} )(\Delta{A_k})(\Delta{A_l})|\psi \rangle } &\!\! {= \langle {A_j}{A_k}{A_l}\rangle \!-\! \langle {A_j}\rangle \langle {A_k}{A_l}\rangle }\\
{} &\!\! {\phantom{=} -\! \langle {A_k}\rangle \langle {A_j}{A_l}\rangle \!-\! \langle {A_l}\rangle \langle {A_j}{A_k}\rangle}\\
{} &\!\! {\phantom{=} + 2\langle {A_j}\rangle \langle {A_k}\rangle \langle {A_l}\rangle,}
\end{array}\!
\end{Equation}
which raises the question; would we group the second two deviations with the ket and the first deviation with the bra, or some other grouping? However, this \textit{does} give us a clear generalization of multi-operator covariance, which we might call the \textit{multivariance},
\begin{Equation}                      {81}
\sigma _{{A_1}, \ldots ,{A_M}}^M \equiv {\mathop{\rm cov}} ({A_1}, \ldots ,{A_M}) \equiv \left\langle {\prod\limits_{j = 1}^M {({A_j} - \langle {A_j}\rangle )} } \right\rangle ,
\end{Equation}
simplifying to \Eq{8} with $\Delta A_j \equiv {A_j} - \langle {A_j}\rangle$, with analogous generalizations of \Eq{79} and \Eq{80} for each $M$. (Although multivariance is not symmetric [invariant under permutations of its arguments], it can be used to construct a \textit{symmetric multivariance}, which we discuss in \Sec{V}.  Here, we focus on \Eq{81} since it can reveal a finer level of detail by describing operators in a certain order.)

But now we see that there is more than one way to express this quantity as an overlap (even for $M = 2$). For example, for $M = 3$ Hermitian operators, if we define
\begin{Equation}                      {82}
\begin{array}{*{20}{l}}
{|{\psi _j}\rangle }&\!\!{ \equiv ({A_j} - \langle {A_j}\rangle )|\psi \rangle ,}\\
{|{\psi _{j,k}}\rangle }&\!\!{ \equiv ({A_j} - \langle {A_j}\rangle )({A_k} - \langle {A_k}\rangle )|\psi \rangle ,}\\
{|{\psi _{j,k,l}}\rangle }&\!\!{ \equiv ({A_j} - \langle {A_j}\rangle )({A_k} - \langle {A_k}\rangle )({A_l} - \langle {A_l}\rangle )|\psi \rangle, }
\end{array}
\end{Equation}
then the multivariance of ${A_1},{A_2},{A_3}$ can be expressed as
\begin{Equation}                      {83}
\begin{array}{*{20}{l}}
{\sigma _{{A_1},{A_2},{A_3}}^3}&\!\!{ =\langle \psi|\psi_{1,2,3} \rangle} &\!\!{= (\langle \psi_{1,2,3}|\psi \rangle)^*}\\
{}&\!\!{ =\langle {\psi _1}|{\psi _{2,3}}\rangle} &\!\!{= (\langle {\psi _{2,3}}|{\psi _{1}}\rangle)^*}\\
{}&\!\!{ =\langle {\psi _{2,1}}|{\psi _{3}}\rangle} &\!\!{=(\langle {\psi _{3}}|{\psi _{2,1}}\rangle)^* }\\
{}&\!\!{ =\langle {\psi _{3,2,1}}|\psi \rangle} &\!\!{= (\langle {\psi }|\psi_{3,2,1} \rangle)^*}\\
\end{array}
\end{Equation}
which means that any linear combination of these where the coefficients sum to 1 would also yield the same quantity.  For the purposes of uncertainty relations, we are only interested in the magnitude of this quantity, so that narrows things down a bit as
\begin{Equation}                      {84}
\begin{array}{*{20}{l}}
{|\sigma _{{A_1},{A_2},{A_3}}^3|}&\!\!{ =|\langle \psi|\psi_{1,2,3} \rangle|} \\
{}&\!\!{ =|\langle {\psi _1}|{\psi _{2,3}}\rangle|}\\
{}&\!\!{ =|\langle {\psi _{2,1}}|{\psi _{3}}\rangle|} \\
{}&\!\!{ =|\langle {\psi _{3,2,1}}|\psi \rangle|.}\\
\end{array}
\end{Equation}
Then, applying the quantum-state CS inequality from \Eq{48} for two operators to each term in \Eq{84} yields
\begin{Equation}                      {85}
\begin{array}{*{20}{r}}
{\sqrt {\langle \psi |\psi \rangle } \sqrt {\langle {\psi _{1,2,3}}|{\psi _{1,2,3}}\rangle } }&\!\!{ \ge |\sigma _{{A_1},{A_2},{A_3}}^3|,}\\
{\sqrt {\langle {\psi _1}|{\psi _1}\rangle } \sqrt {\langle {\psi _{2,3}}|{\psi _{2,3}}\rangle } }&\!\!{ \ge |\sigma _{{A_1},{A_2},{A_3}}^3|,}\\
{\sqrt {\langle {\psi _{2,1}}|{\psi _{2,1}}\rangle } \sqrt {\langle {\psi _3}|{\psi _3}\rangle } }&\!\!{ \ge |\sigma _{{A_1},{A_2},{A_3}}^3|,}\\
{ \sqrt {\langle {\psi _{3,2,1}}|{\psi _{3,2,1}}\rangle }\sqrt {\langle \psi |\psi \rangle } }&\!\!{ \ge |\sigma _{{A_1},{A_2},{A_3}}^3|,}
\end{array}
\end{Equation}
and we can get an infinite number of other relations from convex combinations of them.  Converting the kets in \Eq{85} to vectors as $|\psi_{j_1,\ldots,j_q}\rangle\to\mathbf{a}_{j_1,\ldots,j_q}$ [as defined above \Eq{7}] and $|\psi\rangle\to\mathbf{a}$ then yields \Eq{10}.

Then, looking at the overlaps in \Eq{85} we see that they can be expressed as multivariances.  For example,
\begin{Equation}                      {86}
\begin{array}{*{20}{l}}
{\sqrt {\langle {\psi _{2,3}}|{\psi _{2,3}}\rangle } }&\!\!{ = \sqrt {\langle \psi |{{[(\Delta {A_2})(\Delta {A_3})]}^\dag }(\Delta {A_2})(\Delta {A_3})|\psi \rangle } }\\
{}&\!\!{ = \sqrt {\langle (\Delta {A_3})(\Delta {A_2})(\Delta {A_2})(\Delta {A_3})\rangle } }\\
{}&\!\!{ = \sqrt {\sigma _{{A_3},{A_2},{A_2},{A_3}}^4}, }
\end{array}
\end{Equation}
and continuing this way lets us rewrite \Eq{85} as
\begin{Equation}                      {87}
\begin{array}{*{20}{r}}
{\sqrt {\langle \psi |\psi \rangle } \sqrt {\sigma _{{A_3},{A_2},{A_1},{A_1},{A_2},{A_3}}^6} }&\!\!{ \ge |\sigma _{{A_1},{A_2},{A_3}}^3|,}\\
{{\sigma _{{A_1}}}\sqrt {\sigma _{{A_3},{A_2},{A_2},{A_3}}^4} }&\!\!{ \ge |\sigma _{{A_1},{A_2},{A_3}}^3|,}\\
{\sqrt {\sigma _{{A_1},{A_2},{A_2},{A_1}}^4}\sigma_{A_3} }&\!\!{ \ge |\sigma _{{A_1},{A_2},{A_3}}^3|,}\\
{ \sqrt {\sigma _{{A_1},{A_2},{A_3},{A_3},{A_2},{A_1}}^6}\sqrt {\langle \psi |\psi \rangle } }&\!\!{ \ge |\sigma _{{A_1},{A_2},{A_3}}^3|,}
\end{array}
\end{Equation}
which is the quantum version of \Eq{10} for Hermitian operators and not-necessarily-normalized states.

Similarly, for $M=2$, the resulting relations would be
\begin{Equation}                      {88}
\scalemath{0.93}{\begin{array}{*{20}{r}}
{\sqrt {\langle \psi |\psi \rangle } \sqrt {\sigma _{{A_2},{A_1},{A_1},{A_2}}^4} }&\!\!{ \ge |\sigma _{{A_1},{A_2}}^2|,} &{}\\
{{\sigma _{{A_1}}}{\sigma _{{A_2}}}}&\!\!{ \ge|\sigma _{{A_1},{A_2}}^2|\phantom{,}} & {({\text{Schr{\"o}dinger's relation}}),}\\
{ \sqrt {\sigma _{{A_1},{A_2},{A_2},{A_1}}^4}\sqrt {\langle \psi |\psi \rangle } }&\!\!{ \ge |\sigma _{{A_1},{A_2}}^2|,} &{}
\end{array}}
\end{Equation}
which is the quantum version of \Eq{9}, again with the understanding that we can make any convex combination of these. Note that \Eq{9} and \Eq{10} were derived from the quantum versions by reversing the substitution in \Eq{44}.

Generalizing this, \Eq{81} can be expanded as
\begin{Equation}                      {89}
\sigma _{{A_1}, \ldots ,{A_M}}^M \equiv \langle \psi |(\Delta {A_1}) \cdots (\Delta {A_p})(\Delta {A_{p + 1}}) \cdots (\Delta {A_M})|\psi \rangle ,
\end{Equation}
for any integer $p \in 0, \ldots ,M$, where if $p=0$, that means to partition the product of deviations as $I \times [(\Delta {A_1}) \cdots (\Delta {A_M})]$, and if $p=M$ then partition them as $[(\Delta {A_1}) \cdots (\Delta {A_M})]\times I$.  Then, defining
\begin{Equation}                      {90}
\begin{array}{*{20}{l}}
{|{\psi _{{j_1}, \ldots ,{j_q}}}\rangle }&\!\!{ \equiv (\Delta {A_{{j_1}}}) \cdots (\Delta {A_{{j_q}}})|\psi \rangle, }\\
{\langle {\psi _{{j_1}, \ldots ,{j_q}}}|}&\!\!{ = \langle \psi |(\Delta {A_{{j_q}}}) \cdots (\Delta {A_{{j_1}}}),}
\end{array}
\end{Equation}
we see that the $p$ to $p+1$ transition in the product in \Eq{89} partitions the expectation value into an overlap between two of these not-necessarily normalized states as
\begin{Equation}                      {91}
\sigma _{{A_1}, \ldots ,{A_M}}^M \equiv \langle {\psi _{p, \ldots ,1}}|{\psi _{p + 1, \ldots ,M}}\rangle ,
\end{Equation}
where if $p =0$, we let $\langle {\psi _{0, \ldots ,1}}|{\psi _{0 + 1, \ldots ,M}}\rangle  \equiv \langle \psi |{\psi _{1, \ldots ,M}}\rangle $ and if $p=M$,  $\langle {\psi _{M, \ldots ,1}}|{\psi _{M+1, \ldots ,M}}\rangle  \equiv \langle {\psi _{M, \ldots ,1}}|\psi \rangle $. Then, applying the regular CS inequality to \Eq{91} gives
\begin{Equation}                      {92}
\scalemath{0.86}{\sqrt {\langle {\psi _{p, \ldots ,1}}|{\psi _{p, \ldots ,1}}\rangle } \sqrt {\langle {\psi _{p + 1, \ldots ,M}}|{\psi _{p + 1, \ldots ,M}}\rangle }  \ge |\langle {\psi _{p, \ldots ,1}}|{\psi _{p + 1, \ldots ,M}}\rangle |},
\end{Equation}
which, putting \Eq{91} into the right side of \Eq{92} as
\begin{Equation}                      {93}
\sqrt {\langle {\psi _{p, \ldots ,1}}|{\psi _{p, \ldots ,1}}\rangle } \sqrt {\langle {\psi _{p + 1, \ldots ,M}}|{\psi _{p + 1, \ldots ,M}}\rangle }  \ge |\sigma _{{A_1}, \ldots ,{A_M}}^M  |,
\end{Equation}
proves line 1 of \Eq{13}. Or, using the facts that
\begin{Equation}                      {94}
\scalemath{0.76}{\begin{array}{*{20}{l}}
{\langle {\psi _{p, \ldots ,1}}|{\psi _{p, \ldots ,1}}\rangle }&\!\!{ \equiv \langle \psi |{{[(\Delta {A_{p}}) \cdots (\Delta {A_{1}})]}^\dag }(\Delta {A_{p}}) \cdots (\Delta {A_{1}})|\psi \rangle }\\
{}&\!\!{ = \langle \psi |(\Delta {A_{1}}) \cdots (\Delta {A_{p}})(\Delta {A_{p}}) \cdots (\Delta {A_{1}})|\psi \rangle }\\
{}&\!\!{ = \sigma _{{A_{1}, \ldots ,A_{p},A_{p}, \ldots ,A_{1}}}^{2p},}
\end{array}}
\end{Equation}
and similarly,
\begin{Equation}                      {95}
\scalemath{0.76}{\begin{array}{*{20}{l}}
{\langle {\psi _{p+1, \ldots ,M}}|{\psi _{p+1, \ldots ,M}}\rangle }&\!\!{ \equiv \langle \psi |{{[(\Delta {A_{p+1}}) \cdots (\Delta {A_{M}})]}^\dag }(\Delta {A_{p+1}}) \cdots (\Delta {A_{M}})|\psi \rangle }\\
{}&\!\!{ = \langle \psi |(\Delta {A_{M}}) \cdots (\Delta {A_{p+1}})(\Delta {A_{p+1}}) \cdots (\Delta {A_{M}})|\psi \rangle }\\
{}&\!\!{ = \sigma _{{A_{M}, \ldots ,A_{p+1},A_{p+1}, \ldots ,A_{M}}}^{2(M-p)},}
\end{array}}
\end{Equation}
lets us write \Eq{93} in terms of multivariances as
\begin{Equation}                      {96}
\scalemath{0.84}{\sqrt {\sigma _{{A_{1}, \ldots ,A_{p},A_{p}, \ldots ,A_{1}}}^{2p}} \sqrt {\sigma _{{A_{M}, \ldots ,A_{{p + 1}},A_{{p + 1}}, \ldots ,A_{M}}}^{2(M - p)}}  \ge |\sigma _{{A_{1}}, \ldots ,{A_{M}}}^M|},
\end{Equation}
which proves line 2 of \Eq{13}. Our asymmetrical definition of multivariance gives us a fine degree of specificity and lets us use multivariances on both sides of the inequality. See \Sec{V} for its extension to symmetric multivariance.
\section{\label{sec:IV}Multi-Operator Squeezing}
\TOCSecTarget{IV}{-39pt}
Our balanced multi-operator uncertainty relations show that instead of one quantity being uncertain when the other is certain, we instead get multi-bin generalizations of that, with multiple possible combinations.

This suggests a new way to define \textit{multi-operator squeezing}, a topic that already has a rich history established in the related idea of \textit{multimode squeezing} \cite{BaKn,YuPo,BKn2,CoMM,GeKn}, but which may benefit from our new definitions here.

For instance, the traditional definition of a squeezed state \cite{MoG1,MoG2,Stol,LuCo,Yuen,GeKn} is that given some two-operator uncertainty relation such as the Robertson version of \Eq{1}, if we square it to get a variance relation as
\begin{Equation}                      {97}
{\textstyle \sigma_A^2\sigma_B^2 \ge \frac{{|\langle [A,B]\rangle {|^2}}}{4}},
\end{Equation}
then a state is called \textit{squeezed} (wrt that uncertainty relation and these operators) if
\begin{Equation}                      {98}
{\textstyle \sigma _A^2 < \frac{{|\langle [A,B]\rangle |}}{2}\quad \text{or}\quad \sigma _B^2 < \frac{{|\langle [A,B]\rangle |}}{2}},
\end{Equation}
which is significant because both uncertainties cannot be below the values on the right in \Eq{98} simultaneously.

As we know, this is relevant to coherent states \cite{Gla1,Gla2,GeKn}, which achieve equalities for both relations in \Eq{98} simultaneously, thus saturating \Eq{97}.  And yet, this definition is somewhat \textit{arbitrary}, because it is based on a simplified version of the Schr{\"o}dinger relation (which would have been tighter), and also because the choice to use the square root of the right side of \Eq{97} in the right sides of \Eq{98} is arbitrary even though it is egalitarian [e.g., replacing the right sides of \Eq{98} with the right side of \Eq{97} raised to any positive exponents that add to $1$ would also work].  Therefore, to generalize this, we might reconsider at least some of these choices to allow more options.

From \Eq{23}, the squared generalized uncertainty relation (to get the generalized variance product) for three operators takes the Schr{\"o}dinger form
\begin{Equation}                      {99}
{\widetilde{\sigma}_A^2}{\widetilde{\sigma}_B^2}{\widetilde{\sigma}_C^2} \ge |\widetilde{\sigma}_{A,B}^2||\widetilde{\sigma}_{A,C}^2||\widetilde{\sigma}_{B,C}^2|,
\end{Equation}
noting that from \Eq{26}, \smash{${\widetilde{\sigma}_A^2} = \widetilde{\sigma}_{A,A}^2$}. Abbreviating the right side of \Eq{99} as the \textit{three-operator variance-product bound} \smash{$\beta  \equiv |\widetilde{\sigma}_{A,B}^2||\widetilde{\sigma}_{A,C}^2||\widetilde{\sigma}_{B,C}^2|$} and using egalitarian exponents, \textit{one way} to generalize the notion of squeezing to mean the state is three-operator squeezed is if
\begin{Equation}                      {100}
{\widetilde{\sigma}_A^2} < {\beta ^{1/3}}\quad \text{or}\quad {\widetilde{\sigma}_B^2} < {\beta ^{1/3}}\quad \text{or}\quad {\widetilde{\sigma}_C^2} < {\beta ^{1/3}}.
\end{Equation}
But the extra operator here allows another possibility; we could have \textit{two} variances below this limit, such as
\begin{Equation}                      {101}
\begin{array}{*{20}{l}}
{}&{({{\widetilde{\sigma} }_A^2} < {\beta ^{1/3}}}&{\text{and}}&{{{\widetilde{\sigma} }_B^2} < {\beta ^{1/3}}\;\;\text{s.t.}\;\;\widetilde{\sigma}_A^2\widetilde{\sigma}_B^2\ge |\widetilde{\sigma}_{A,B}^2|)}\\
{\text{or}}&{({{\widetilde{\sigma} }_A^2} < {\beta ^{1/3}}}&{\text{and}}&{{{\widetilde{\sigma} }_C^2} < {\beta ^{1/3}}\;\;\text{s.t.}\;\;\widetilde{\sigma}_A^2\widetilde{\sigma}_C^2\ge |\widetilde{\sigma}_{A,C}^2|)}\\
{\text{or}}&{({{\widetilde{\sigma} }_B^2} < {\beta ^{1/3}}}&{\text{and}}&{{{\widetilde{\sigma} }_C^2} < {\beta ^{1/3}}\;\;\text{s.t.}\;\;\widetilde{\sigma}_B^2\widetilde{\sigma}_C^2\ge |\widetilde{\sigma}_{B,C}^2|),}
\end{array}
\end{Equation}
as long as the third one is big enough to maintain ${\widetilde{\sigma}_A^2}{\widetilde{\sigma}_B^2}{\widetilde{\sigma}_C^2} \ge \beta$ which is \Eq{99}. Thus, \Eq{100} is a more loose definition of three-operator squeezing that we might call ``$1/3$ operator squeezing,'' (or just ``$1/3$ squeezing'') and \Eq{101} is a more stringent definition that we might call ``$2/3$ squeezing.''
 
For $1/3$ squeezing, the two operators whose variances are not squeezed can obey a two-operator inequality of their own, as long as all three still obey \Eq{99}, and the two-operator uncertainty relations that share the squeezed variance are still obeyed.  Thus, $1/3$ operator squeezing has the latitude to permit ordinary two-operator squeezing ($1/2$ squeezing in this nomenclature) as well.

On the other hand, the $2/3$ squeezing in \Eq{101} is \textit{too squeezed} to permit two-operator squeezing, and can be thought of as genuine three-operator squeezing.

In general, $q/M$ \textit{squeezing} is possible for $q \in 1, \ldots ,M - 1$, and if the lower limit of its generalized variance product (with respect to a particular $M$-operator uncertainty relation) is $\beta$, then the upper limit of generalized variance for an operator being squeezed is \smash{$\beta^{1/M}$}, and there are \smash{$\binom{M}{q}$} ways to achieve that squeezing.  Thus, using the balanced any-operator uncertainty relation from \Eq{23} for $M$ operators, we say a state is $q/M$ \textit{squeezed} if any particular subset \smash{$\{ A_{j_1}, \ldots ,A_{j_q}\} $} of those operators obeys
\begin{Equation}                      {102}
\setlength\fboxsep{4pt}   
\setlength\fboxrule{0.5pt}
\fbox{$\begin{array}{*{20}{l}}
{{{\widetilde{\sigma} }_{{A_{{j_1}}}}^2}\!\!< {\beta ^{1/M}} \ldots \,\,\text{and}\,\, \ldots\; {{\widetilde{\sigma} }_{{A_{{j_q}}}}^2}\!\! < {\beta ^{1/M}}}\\
{\text{s.t. all uncertainty relations are still obeyed}}
\end{array}$}\;,
\end{Equation}
where \smash{$\beta^{1/M}$} is the $q/M$ \textit{squeezing threshold}, and by squaring \Eq{23}, the \textit{$M\!$-operator variance-product bound} is
{\\\vspace{-17pt}}%
\begin{Equation}                      {103}
\beta  \equiv {\left( {\scalemath{0.85}{\prod\limits_{j = 1}^{M - 1} {\prod\limits_{k = j + 1}^M {|\widetilde{\sigma}_{{A_j},{A_k}}^2|} }} } \right)^{\!\!2/(M - 1)}},
\end{Equation}
{~\vspace{-9pt}}\\%
and the ``uncertainty relations still obeyed'' are all balanced any-operator uncertainty relations from \Eq{23} for this $M$ or smaller $M$, where the uncertainty type should be the same (i.e., if using the Robertson version for $M$ operators, use that for all relations of smaller $M$ as well).

\begin{widetext}~
{\\\vspace{-30pt}}
\begin{table}[H]
\scalebox{0.893}{\begin{threeparttable}[b]
\caption{\label{tab:1}Types of multi-operator squeezing for $M=3$ operators where \smash{$\beta  \equiv |\widetilde{\sigma}_{A,B}^2||\widetilde{\sigma} _{A,C}^2||\widetilde{\sigma}_{B,C}^2|$} and we use specific factors $a,b,c > 1$ to keep track of exactly how each quantity must change to obey the multi-operator uncertainty relation(s) involved. For example, an entry in the first column such as \smash{$=\frac{{{\beta ^{1/3}}}}{a}<\beta^{1/3}$} is to be read as ``\smash{${\widetilde{\sigma}_A^2} = \frac{\beta ^{1/3}}{a}$} which is less than \smash{$\beta ^{1/3}$} by a factor of \smash{$\frac{1}{a}$},'' where it is understood that $a$ itself is not part of the definition of squeezing but rather $a,b,c$ must be chosen such that all uncertainty relations in a given row are obeyed.}
{\renewcommand{\arraystretch}{1.7}
\renewcommand{\tabcolsep}{0.2cm}
\begin{tabular}{|l|l|l|l|l|l|l|}
\hline
{$\hsp{29}\widetilde{\sigma}_A^2$} & {$\hsp{29}\widetilde{\sigma}_B^2$} & {$\hsp{29}\widetilde{\sigma}_C^2$} & {$\hsp{25}\widetilde{\sigma}_A^2\widetilde{\sigma}_B^2$} &{$\hsp{25}\widetilde{\sigma}_A^2\widetilde{\sigma}_C^2$} & {$\hsp{25}\widetilde{\sigma}_B^2\widetilde{\sigma}_C^2$} & {\hsp{5.5}Squeezing Type} \\[0.5mm]
\hline
$=\frac{\beta^{1/3}}{a}<\beta^{1/3}$ & $=\frac{\beta^{1/3}}{b}<\beta^{1/3}$ & $=\frac{\beta^{1/3}}{c}<\beta^{1/3}$ & {\hsp{28}N/A} & {\hsp{28}N/A} & {\hsp{28}N/A} & {\!\!\!\! ``$3/3$'' \!{\small (impossible)}\!\!\! } \\[0.5mm]
\hline
$=\frac{\beta^{1/3}}{a}<\beta^{1/3}$ & $=\frac{\beta^{1/3}}{b}<\beta^{1/3}$ & $= ab\beta ^{1/3}> \beta^{1/3}$ & $=\frac{{{\beta ^{2/3}}}}{{ab}}\ge |\widetilde{\sigma}_{A,B}^2|$ &$ =b\beta^{2/3}\ge |\widetilde{\sigma}_{A,C}^2|$ & $= a\beta^{2/3}\geq|\widetilde{\sigma}_{B,C}^2|$ & {\hsp{1.5}$2/3;$ $\beta\geq 1$} \\[0.5mm]
\hline
$=\frac{\beta^{1/3}}{a}<\beta^{1/3}$ & $=ac\beta ^{1/3}>\beta ^{1/3}$ & $=\frac{\beta^{1/3}}{c}<\beta^{1/3}$ & $=c\beta ^{2/3}\ge |\widetilde{\sigma}_{A,B}^2|$ &$=\frac{{{\beta ^{2/3}}}}{{ac}}\ge |\widetilde{\sigma}_{A,C}^2|$ & $=a\beta ^{2/3}\ge |\widetilde{\sigma}_{B,C}^2|$ & {\hsp{1.5}$2/3;$ $\beta\geq 1$} \\[0.5mm]
\hline
$= bc\beta ^{1/3}>\beta ^{1/3}$ & $=\frac{\beta^{1/3}}{b}<\beta^{1/3}$ & $=\frac{\beta^{1/3}}{c}<\beta^{1/3}$ & $=c\beta ^{2/3}\ge |\widetilde{\sigma}_{A,B}^2|$ &$=b\beta ^{2/3}\ge |\widetilde{\sigma}_{A,C}^2|$ & $=\frac{{{\beta ^{2/3}}}}{{bc}}\ge |\widetilde{\sigma}_{B,C}^2|$ & {\hsp{1.5}$2/3;$ $\beta\geq 1$} \\[0.5mm]
\hline
$=\frac{\beta^{1/3}}{bc}<\beta^{1/3}$ & $=b\beta^{1/3}>\beta^{1/3}$ & $=c\beta^{1/3}>\beta^{1/3}$ & $=\frac{\beta^{2/3}}{c}\ge |\widetilde{\sigma}_{A,B}^2|$ &$=\frac{\beta^{2/3}}{b}\ge |\widetilde{\sigma}_{A,C}^2|$ & $=bc{\beta ^{2/3}} \ge |\widetilde{\sigma}_{B,C}^2|$ & {\hsp{1.5}$1/3;\!\!$ \smash{$\shiftmath{1.7pt}{\phantom{|}_{|\widetilde{\sigma}_{B,C}^2|\geq bc\beta^{-\!1/3}}^{\beta\geq 1}}\!$}} \\[0.5mm]
\hline
$=a\beta^{1/3}>\beta^{1/3}$ & $=\frac{\beta^{1/3}}{ac}<\beta^{1/3}$ & $=c\beta^{1/3}>\beta^{1/3}$ & $=\frac{\beta^{2/3}}{c}\ge |\widetilde{\sigma}_{A,B}^2|$ &$= ac{\beta ^{2/3}} \ge |\widetilde{\sigma}_{A,C}^2|$ & $=\frac{\beta^{2/3}}{a}\ge |\widetilde{\sigma}_{B,C}^2|$ & {\hsp{1.5}$1/3;\!\!$ \smash{$\shiftmath{1.7pt}{\phantom{|}_{|\widetilde{\sigma}_{A,C}^2|\geq ac\beta^{-\!1/3}}^{\beta\geq 1}}\!$}} \\[0.5mm]
\hline
$=a\beta^{1/3}>\beta^{1/3}$ & $=b\beta^{1/3}>\beta^{1/3}$ & $=\frac{\beta^{1/3}}{ab}<\beta^{1/3}$ & $= ab{\beta ^{2/3}} \ge |\widetilde{\sigma}_{A,B}^2|$ &$=\frac{\beta^{2/3}}{b}\ge |\widetilde{\sigma}_{A,C}^2|$ & $=\frac{\beta^{2/3}}{a}\ge |\widetilde{\sigma}_{B,C}^2|$ & {\hsp{1.5}$1/3;\!\!$ \smash{$\shiftmath{1.7pt}{\phantom{|}_{|\widetilde{\sigma}_{A,B}^2|\geq ab\beta^{-\!1/3}}^{\beta\geq 1}}\!$}} \\[0.5mm]
\hline
$= a\beta^{1/3}\geq\beta^{1/3}$ & $= b\beta^{1/3}\geq\beta^{1/3}$ & $= c\beta^{1/3}\geq\beta^{1/3}$ & $= ab{\beta ^{2/3}} \ge |\widetilde{\sigma}_{A,B}^2|$ &$= ac{\beta ^{2/3}} \ge |\widetilde{\sigma}_{A,C}^2|$ & $= bc{\beta ^{2/3}} \ge |\widetilde{\sigma}_{B,C}^2|$ & {\!\!\!\! ``$0/3$'' \!{\small (none)} ${\scriptstyle a,b,c\geq 1}\!$} \\
\hline
\end{tabular}}
\end{threeparttable}}
\end{table}~
{\\\vspace{-28pt}}
\end{widetext}

As a three-operator example, \Table{1} shows all the combinations for which multi-operator squeezing can take place by our definition above. Note that the type of uncertainty relation used to define this can lead to different results for a given state and set of operators, but that is true of the original notion of squeezing as well.

\begin{figure}[H]
\centering
\includegraphics[width=1.00\linewidth]{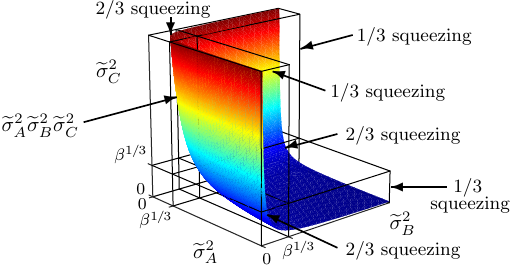}
\vspace{-14pt}
\caption[]{Schematic visualization of three-operator squeezing for constant $\beta$. The large flat square regions of thickness $\beta^{1/3}$ and larger extent in their two other directions contain $1/3$ squeezing since only one of the three operators is squeezed in those regions. The thin long rectangular-prism regions where those flat squares intersect contain $2/3$ squeezing.}
\label{fig:6}
\end{figure}

\Figure{6} visualizes three-operator squeezing for an example where $\beta$ is constant. The cube of sides $\beta^{1/3}$ near the origin is the cube of impossibility (we might call that ``$3/3$ squeezing'' except no state could achieve it for any operators). The largest region where all variances are simultaneously at least $\beta^{1/3}$ could be called ``$0/3$ squeezing'' to mean ``no squeezing,'' which includes $\beta =0$.

A full exploration of multi-operator squeezing is beyond the scope of this paper, but due to the many applications of previously defined notions of squeezing, this is likely to be an interesting topic for future research.
\section{\label{sec:V}Conclusions}
\TOCSecTarget{V}{-39pt}
In this paper, we have investigated several new quantum uncertainty relations motivated by several new types of Cauchy-Schwarz inequalities for multiple vectors.

These new results generate a wealth of new perspectives for understanding relationships between different observables for a given state or family of states, as seen in our various diagrams.

While most other proposed methods of generalizing the uncertainty relations focus on improving the \textit{tightness} of their inequalities, we focus instead on \textit{conceptual simplicity} of inequalities.

The reason for this is that since the CS inequalities and their multi-operator generalizations essentially come from ignoring the cosine factor of each overlap, then leaving that in for each overlap would give the tightest possible bound, which is always simply the computation of the product of uncertainties explicitly.

This is less trivial than it may seem; the fact that any novel results appear at all in the quantum uncertainty relations comes from the overlaps themselves.  The inequalities merely come from ignoring factors from those overlaps, such as we just described for the CS inequality.

Therefore, there are not many advantages to using inequalities except to highlight the wave behavior that arises from the quantum nature of the systems being described, i.e., when position is known, momentum is uncertain, and vice versa. However, the inequality form of uncertainty relations does lead to important concepts such as squeezing which has applications in metrology, and we have shown that the simplicity of our multi-operator uncertainty relations makes multi-operator squeezing tractable. Thus, the main benefit of using inequalities is the simplicity they can bring to comparisons of multiple observables, not their tightness [because we can always just compute the quantities themselves for the tightest relation, as in (22) and (31)].

For instance, if \textit{tightness} is the goal, the Robertson and Schr{\"o}dinger relations in \Eq{1} could simply be amended to
\begin{Equation}                      {104}
{\sigma _A}{\sigma _B} = \sqrt {\langle {A^2}\rangle  - {\langle A\rangle^2}} \sqrt {\langle {B^2}\rangle  - {\langle B\rangle^2}} ,
\end{Equation}
and no information would be lost; the right side is the tightest bound, and contains all novel quantum effects relevant to these quantities.  The actual Robertson and Schr{\"o}dinger relations sacrifice some information by using the CS inequalities.

It should also be noted that although this is all very interesting, the covariance and its normalized form as the \textit{Pearson correlation} \cite{Galt,Pear,Devo} are only measures of \textit{linear} correlation, and similar restrictions apply to our generalization of it as the multivariance in \Eq{8}.

As a powerful alternative, a measure of all possible nonlocal correlations in a state is the \textit{correlance} \cite{HCor}, although it is a state-centric measure rather than an operator-centric measure. However, that paper explains how to convert any classical multivariable data set into an equivalent diagonal quantum state to be used with the measure to detect its correlations.  (In the case of a quantum system, one could use quantum state tomography \cite{JKMW,HedD} with a tomographically complete set of generally nondiagonal observables to construct an estimator of the quantum density operator, and then the correlance would detect any correlations in that state.)  The correlance is not limited to linear correlations, it can handle any number of discrete operators, and it is explicitly computable for mixed states as well as pure states.

The only caveat is that correlance requires a \textit{multipartite} system because it measures all \textit{nonlocal} correlations, whereas the relations in the present paper can work on unipartite systems, i.e. systems with no fundamental coincidence outcomes and therefore no nonlocal correlations, such as single qubits. Nevertheless, correlance provides an extremely useful alternative to covariance and the Pearson correlation, since it can detect correlations that they cannot.

Regarding the multivariance from \Eq{8}, \Sec{II.E}, and \Sec{III.D}, we presented it in asymmetrical form because covariance itself is not symmetric yet is part of the standard uncertainty relations, but also because its asymmetric form gives us a high degree of specificity for investigating multiple operators.  Nevertheless, we offer here a definition of symmetric multivariance, since that may be more desirable in some applications, focusing our discussion on Hermitian operators for simplicity.

Recall\hsp{-1.5} that\hsp{-1.5} for\hsp{-1.5} two\hsp{-1.5} Hermitian\hsp{-1.5} operators,\hsp{-1.5} the\hsp{-1.5} covariance,
\begin{Equation}                      {105}
\sigma _{{A_1},{A_2}}^2 \equiv \langle (\Delta {A_1})(\Delta {A_2})\rangle  = \langle {A_1}{A_2}\rangle  - \langle {A_1}\rangle \langle {A_2}\rangle ,
\end{Equation}
is not symmetric since \smash{$\langle {A_2}{A_1}\rangle =\langle {A_2^\dag}{A_1^\dag}\rangle  = \langle {({A_1}{A_2})^\dag }\rangle =$} \smash{$ \langle {A_1}{A_2}\rangle^{*}$}.  But we can symmetrize \Eq{105} by making it invariant under permutations of its arguments as
\begin{Equation}                      {106}
{\textstyle \hat{\sigma}_{{A_1},{A_2}}^2 \equiv \frac{1}{2}(\sigma _{{A_1},{A_2}}^2 + \sigma _{{A_2},{A_1}}^2) = \frac{{\langle \{ {A_1},{A_2}\} \rangle }}{2} - \langle {A_1}\rangle \langle {A_2}\rangle},
\end{Equation}
which is not the only way to achieve symmetry, but this form is already in wide usage, such as in \cite{Buon,HeCl}.

For three Hermitian operators, the symmetric multivariance would be
\begin{Equation}                      {107}
\begin{array}{*{20}{l}}
{\hat \sigma _{{A_1},{A_2},{A_3}}^3}&\!\!{ \equiv \frac{1}{6}(\phantom{+} \sigma _{{A_1},{A_2},{A_3}}^3 + \sigma _{{A_1},{A_3},{A_2}}^3 + \sigma _{{A_2},{A_3},{A_1}}^3 }\\
{}&\!\!{\phantom{\equiv \frac{1}{6}}\hsp{-0.1} + \sigma _{{A_2},{A_1},{A_3}}^3 + \sigma _{{A_3},{A_1},{A_2}}^3 + \sigma _{{A_3},{A_2},{A_1}}^3)}\\
{}&\!\!{ = \frac{{\langle {A_1}\{ {A_2},{A_3}\}  + {A_2}\{ {A_3},{A_1}\}  + {A_3}\{ {A_1},{A_2}\} \rangle }}{6}}\\
{}&\!\!{\phantom{=} - \frac{{\langle {A_1}\rangle \langle \{ {A_2},{A_3}\} \rangle  + \langle {A_2}\rangle \langle \{ {A_3},{A_1}\} \rangle  + \langle {A_3}\rangle \langle \{ {A_1},{A_2}\} \rangle }}{2} }\\
{}&\!\!{\phantom{=} + 2\langle {A_1}\rangle \langle {A_2}\rangle \langle {A_3}\rangle,}
\end{array}
\end{Equation}
where notice that the special form in line 2 is a jump in complexity beyond the special form in \Eq{106}.

However, the special forms are not as important as the definition, so we can define the \textit{symmetric multivariance for $M$ Hermitian operators} as
\begin{Equation}                      {108}
\setlength\fboxsep{4pt}   
\setlength\fboxrule{0.5pt}
\fbox{$\hat \sigma _{{A_1}, \ldots ,{A_M}}^M \equiv \frac{1}{{M!}}\sum\limits_{q = 1}^{M!} {\sigma _{{P_q}({A_1}, \ldots ,{A_M})}^M} $}\;,
\end{Equation}
where ${P_q}({A_1}, \ldots ,{A_M})$ is the $q$th permutation of $\{ {A_1}, \ldots ,{A_M}\} $ in some ordered list of all enumerations of permutations of $\{ {A_1}, \ldots ,{A_M}\} $, such as seen in \Eq{107}.

To get an uncertainty relation for this, first take absolute values in \Eq{108} and use the triangle inequality to get
\begin{Equation}                      {109}
{\textstyle \frac{1}{{M!}}\sum\limits_{q = 1}^{M!} {|\sigma _{{P_q}({A_1}, \ldots ,{A_M})}^M|}  \ge |\hat \sigma _{{A_1}, \ldots ,{A_M}}^M|}.
\end{Equation}
Then, from our results for the asymmetric multivariance, note that the relation in line 2 of \Eq{13} also applies to any permutation as
\begin{Equation}                      {110}
\begin{array}{*{20}{l}}
{\phantom{\times}\sqrt {\sigma _{{A_{{j_{q,1}}}}, \ldots ,{A_{{j_{q,p}}}},{A_{{j_{q,p}}}}, \ldots ,{A_{{j_{q,1}}}}}^{2p}}}&\!\!{}\\[5.5pt]
{\times\sqrt {\sigma _{{A_{{j_{q,M}}}}, \ldots ,{A_{{j_{q,p + 1}}}},{A_{{j_{q,p + 1}}}}, \ldots ,{A_{{j_{q,M}}}}}^{2(M - p)}}}&\!\!{\ge |\sigma _{{A_{{j_{q,1}}}}, \ldots ,{A_{{j_{q,M}}}}}^M|,}
\end{array}
\end{Equation}
where ${A_{{j_{q,1}}}}, \ldots ,{A_{{j_{q,M}}}} \equiv {P_q}({A_1}, \ldots ,{A_M})$.  However, this raises the question: how do we identify which value of $p$ to use when each permutation has $M +1$ different values for $p$? There is no simple answer to this, but the most conceptually consistent way to proceed is to add all of them up with equal weight, using the earlier observation that any convex combination of them is also valid.  Therefore, for a \textit{particular permutation} such as in \Eq{110}, we can add up all $M +1$ partitional variations of that as
\begin{Equation}                      {111}
\scalemath{0.96}{\begin{array}{*{20}{l}}
{{\textstyle \frac{1}{{M + 1}}}\sum\limits_{p = 0}^M {\sqrt {\sigma _{{A_{{j_{q,1}}}}, \ldots ,{A_{{j_{q,p}}}},{A_{{j_{q,p}}}}, \ldots ,{A_{{j_{q,1}}}}}^{2p}} } }&\!\!\!{}\\
{\phantom{{\textstyle \frac{1}{{M + 1}}}} \times\! \sqrt {\sigma _{{A_{{j_{q,M}}}}, \ldots ,{A_{{j_{q,p + 1}}}},{A_{{j_{q,p + 1}}}}, \ldots ,{A_{{j_{q,M}}}}}^{2(M - p)}} }&\!\!\!{ \ge |\sigma _{{P_q}({A_1}, \ldots ,{A_M})}^M|,}
\end{array}}\!
\end{Equation}
where the right side stayed the same because it is the same for all values of $p$ in \Eq{110} and the weights add to $1$. Then, summing \Eq{111} over all permutations and putting that into \Eq{109} gives the \textit{symmetric multivariance uncertainty relation} as
\begin{Equation}                      {112}
\setlength\fboxsep{4pt}   
\setlength\fboxrule{0.5pt}
\fbox{$\scalemath{0.94}{\begin{array}{*{20}{l}}
{\frac{1}{{(M + 1)!}}\sum\limits_{q = 1}^{M!} {\sum\limits_{p = 0}^M {\sqrt {\sigma _{{A_{{j_{q,1}}}}, \ldots ,{A_{{j_{q,p}}}},{A_{{j_{q,p}}}}, \ldots ,{A_{{j_{q,1}}}}}^{2p}} } } }&\!\!\!{}\\
{\phantom{\frac{1}{{(M + 1)!}}} \times\! \sqrt {\sigma _{{A_{{j_{q,M}}}}, \ldots ,{A_{{j_{q,p + 1}}}},{A_{{j_{q,p + 1}}}}, \ldots ,{A_{{j_{q,M}}}}}^{2(M - p)}} }&\!\!\!{ \ge |\hat \sigma _{{A_1}, \ldots ,{A_M}}^M|}
\end{array}}$}
\end{Equation}
where ${A_{{j_{q,1}}}}, \ldots ,{A_{{j_{q,M}}}} \equiv {P_q}({A_1}, \ldots ,{A_M})$ so that $q$ counts over a particular enumeration of the set of indices that distinguishes the product of the $M$ operators.  Needless to say, despite the symmetry of $\hat \sigma _{{A_1}, \ldots ,{A_M}}^M$, for which it was defined, the complicated form on the left is not the most desirable form to work with.

However, the symmetric multivariance \textit{itself} [in \Eq{108}] might be more useful since the tightest inequality is always the quantity itself, in which case the special forms in \Eq{106} and \Eq{107} (or the definition of $\hat \sigma _{{A_1}, \ldots ,{A_M}}^M$ itself) can be those tightest ``inequalities,'' yielding both symmetry and tightness in a single, relatively simple result.

Overall, in terms of simplicity, our generalizations of the CS inequalities and the quantum uncertainty relations they yield have the potential to be more useful than other forms in the literature that focus on tightness.  For instance, as we saw in \Sec{IV}, our balanced multi-operator uncertainty relations yielded a new type of squeezing called \textit{multi-operator squeezing}.  An example that demonstrates that such sets of operators exist is
\begin{Equation}                      {113}
\sigma_{x}\sigma_{p_x}\sigma_{r}\geq \left({\textstyle \tau\frac{\hbar}{2}}\right)^{3/2},
\end{Equation}
where $r\equiv -x-p_{x}$ and \smash{$\tau\equiv \sqrt{\scalemath{0.90}{\frac{4}{3}}}$}, discovered in \cite{KeWe}.  This example also confirms the possibility of a \textit{constant} $q/M$ squeezing threshold, such as \smash{$\beta^{1/3}$} in \Fig{6}.

As with the original definition of squeezing, multi-operator squeezing can exist either for a single mode or for multiple modes (so for instance, it is possible to have multimode multi-operator squeezed states; these notions are neither the same nor mutually exclusive). 

Note that in the rightmost column of \Table{1}, several side conditions emerged because $\beta$ shares factors with the two-operator variance-product bounds, the most common of which is that $\beta\geq 1$. Although this may seem restrictive, it can easily be accommodated by rescaling the operators with an appropriate common factor.

Recalling that we can use the squeezing thresholds [such as the right sides of \Eq{98}] to define a case where the uncertainty relation inequality is \textit{saturated} (achieves equality) in a balanced way, and that this leads to the idea of ``intelligent states'' (coherent states), it is highly likely that a generalization of those states exist for these multi-operator uncertainty relations as well. 

In \cite{KeWe}, it was found that for the particular set of three operators in \Eq{113}, the states that minimize that relation require $\tau$ to be adjusted to $\tau =1$, which allowed them to derive a family of \textit{displaced generalized squeezed states} which included the triple-uncertainty-minimizing states as a special case.  Thus, their example showed that while general multi-operator uncertainty relations exist, the limitations of physical states with such additional properties might impose further constraints that require adjustment of the tightness of a given uncertainty relation.  This would be an interesting area for future research, especially given the technological usefulness of coherent and squeezed states.

Although \cite{KeWe} focused on infinite-dimensional three-operator squeezing, here we have talked about finite-dimensional $M$-operator squeezing, which may not have the same limitations.  Indeed, there already exist several investigations on finite-dimensional squeezing and spin-coherent states \cite{WoKr,Ruzz,GeVZ,MaGD,ViGT}, although none of them have discussed multi-operator generalizations.

In closing, we have now seen that our new CS inequalities and the new quantum uncertainty relations they yield are more than just theoretical curiosities; they lead to more complete understandings of quantum behavior for multiple observables and they also give us clues for how to generalize squeezing to multi-operator squeezing which goes beyond the established idea of multimode squeezing.  Since squeezed states have many important technological applications, this is likely to be an area of fruitful research in the future.  Therefore, the concepts we have presented here extend the work started by Heisenberg and his contemporaries and pave the way for future innovations, both theoretical and technological.
\begin{appendix}
\section{\label{sec:App.A}Adapting These Results to Mixed States}
\TOCAppHeaderTarget{VI}{-39pt}
Here, we show how to adapt all the results of this paper to mixed states, for readers who may be unfamiliar with this elementary procedure. Then we provide a brief proof of why it is valid.

All inequalities in this work involving pure states also apply to mixed states by the following well-known relations.  A mixed state represented by a probability density operator $\rho$ is a convex sum of pure states \smash{$\rho  = \sum\nolimits_j {{p_j}|{\psi _{\{ j\} }}\rangle \langle {\psi _{\{ j\} }}|} $} where \smash{$|{\psi _{\{ j\} }}\rangle $} is the $j$th pure state of a given decomposition of $\rho$ [where we use the curly braces in the subscript as \smash{$|{\psi _{\{ j\} }}\rangle $} so that it is not confused with our earlier definitions that \smash{$|{\psi _j}\rangle  \equiv ({A_j} - \langle {A_j}\rangle )|\psi \rangle $}, and to leave the superscript position open for mode labels to dovetail with notation in our other publications], the $p_j$ are probabilities \smash{$\sum\nolimits_j {{p_j}}  = 1;$ ${p_j} \in [0,1]\,\,\forall j$}, where a pure state \smash{$\rho  = |\psi \rangle \langle \psi |$} can be regarded as a special case where ${p_1} = 1$, and all other probabilities are $0$.  Except for pure states, mixed states have an infinite number of such decompositions.

The expectation value for any operator $A$ for a system in pure state $|\psi \rangle $ is defined as $\langle A\rangle  \equiv \langle \psi |A|\psi \rangle $. Noting that for any set of states $\{ |{\phi _{\{ 1\} }}\rangle , \ldots ,|{\phi _{\{ n\} }}\rangle \} $ (not necessarily decomposition states) that is both orthonormal \smash{$\langle {\phi _{\{ j\} }}|{\phi _{\{ k\} }}\rangle  = {\delta _{j,k}}$} and complete \smash{$\sum\nolimits_{k = 1}^n {|{\phi _{\{ k\} }}\rangle \langle {\phi _{\{ k\} }}|}  = I$}, we can rewrite $\langle A\rangle $ in terms of the pure-state density operator and trace operation as
\begin{Equation}                      {A.1}
{\begin{array}{*{20}{l}}
{\langle A\rangle }&\!\!{ \equiv \langle \psi |A|\psi \rangle  = \langle \psi |A\sum\nolimits_{k = 1}^n {|{\phi _{\{ k\} }}\rangle \langle {\phi _{\{ k\} }}|\psi \rangle } }\\[1pt]
{}&\!\!{= \sum\nolimits_{k = 1}^n {\langle \psi |A|{\phi _{\{ k\} }}\rangle \langle {\phi _{\{ k\} }}|\psi \rangle }}\\[1pt]
{}&\!\!{ = \sum\nolimits_{k = 1}^n {} \langle {\phi _{\{ k\} }}|\psi \rangle \langle \psi |A|{\phi _{\{ k\} }}\rangle }\\[1pt]
{}&\!\!{= \sum\nolimits_{k = 1}^n {} \langle {\phi _{\{ k\} }}|\rho A|{\phi _{\{ k\} }}\rangle\equiv \tr(\rho A),}
\end{array}}
\end{Equation}
where the trace operation, defined on any operator $B$ in the $n$-dimensional Hilbert space as \smash{$\tr(B)\equiv$} \smash{$\sum\nolimits_{k = 1}^n {\langle {\phi _{\{ k\} }}|B|{\phi _{\{ k\} }}\rangle } $}, is independent of the orthonormal complete basis used, and for this pure state, $\rho  = |\psi \rangle \langle \psi |$. Note also that if $A$ is the identity $I$, then \Eq{A.1} gives $\langle I\rangle = \langle \psi |\psi \rangle  = \tr(\rho )$, which equals $1$ for normalized states.

Then, relabeling \smash{$|\psi \rangle $} as $|\psi_{\{j\}} \rangle$ in \Eq{A.1}, multiplying by probability $p_j$, and summing over index $j$ yields \smash{$\sum\nolimits_j {{p_j}\langle {\psi _{\{ j\} }}|A|{\psi _{\{ j\} }}\rangle }=\tr(\sum\nolimits_j{p_j \rho_{\{j\}}A})=\tr(\rho A)$}, which has the same \textit{form} as \Eq{A.1} so it is also written as $\langle A\rangle$ where now \smash{$\rho  \equiv \sum\nolimits_j {{p_j} \rho_{\{j\}}} $} and \smash{$\rho_{\{j\}}\equiv |{\psi _{\{ j\} }}\rangle \langle {\psi _{\{ j\} }}|$}.  Thus, for a \textit{mixed} state $\rho$, we replace the pure $\rho$ in \Eq{A.1} with a density matrix for a mixed state, reusing the same variable but with a different definition as \smash{$\rho  = \sum\nolimits_j {{p_j}|{\psi _{\{ j\} }}\rangle \langle {\psi _{\{ j\} }}|} $} in \Eq{A.1}, which gives
\begin{Equation}                      {A.2}
\begin{array}{*{20}{l}}
{\langle A\rangle }&\!\!{ = \tr(\rho A) = \sum\nolimits_{k = 1}^n {\langle {\phi _{\{ k\} }}|\rho A|{\phi _{\{ k\} }}\rangle } }\\
{}&\!\!{= \sum\nolimits_{k = 1}^n {\langle {\phi _{\{ k\} }}|\sum\nolimits_j {{p_j}|{\psi _{\{ j\} }}\rangle \langle {\psi _{\{ j\} }}|A|{\phi _{\{ k\} }}\rangle } }}\\
{}&\!\!{ = \sum\nolimits_j {\sum\nolimits_{k = 1}^n {{p_j}\langle {\psi _{\{ j\} }}|A|{\phi _{\{ k\} }}\rangle \langle {\phi _{\{ k\} }}|{\psi _{\{ j\} }}\rangle } } }\\[1.5pt]
{}&\!\!{= \sum\nolimits_j {{p_j}\langle {\psi _{\{ j\} }}|A\sum\nolimits_{k = 1}^n {|{\phi _{\{ k\} }}\rangle \langle {\phi _{\{ k\} }}|{\psi _{\{ j\} }}\rangle } }}\\
{}&\!\!{ = \sum\nolimits_j {{p_j}\langle {\psi _{\{ j\} }}|A|{\psi _{\{ j\} }}\rangle }\equiv \sum\nolimits_j {{p_j}{{\langle A\rangle }_j}},  }\\
\end{array}
\end{Equation}
which shows that for a mixed state, the expectation value of any operator is a convex sum of expectation values \smash{${\langle A\rangle _j} \equiv \langle {\psi _{\{ j\} }}|A|{\psi _{\{ j\} }}\rangle $} of that operator for each pure state of a given decomposition.

This same procedure is then used in every quantity that involves an expectation value, which in this work are all generalized and regular covariances and variances, and even the multivariance.  For instance, we give the mixed-state versions of some key quantities below, in order of their appearance in the text for all the quantum results (where as always in this work, all products grow from left to right, and mixed-state versions of the complex vector results follow analogously to these so we omit them here), which will let all the results of this paper be written for mixed states.

The \textit{multivariance} for mixed states (with Hermitian inputs) becomes
\begin{Equation}                      {A.3}
{\sigma _{{A_1}, \ldots ,{A_M}}^M \equiv {\mathop{\rm cov}} ({A_1}, \ldots ,{A_M}) \equiv \tr\left( {\rho \prod\limits_{j = 1}^M {(\Delta {A_j})} } \right) \in \mathbb{C}.}
\end{Equation}
The \textit{covariance}, \textit{variance}, and \textit{standard deviation} for mixed states are
\begin{Equation}                      {A.4}
{\begin{array}{*{20}{l}}
{\sigma _{{A_j},{A_k}}^2}&\!\!{\equiv \langle {\psi _j}|{\psi _k}\rangle  \equiv {\mathop{\rm cov}} ({A_j},{A_k}) \equiv \tr[\rho (\Delta {A_j})(\Delta {A_k})]}\\
{}&\!\!{=\tr[\rho {A_j}{A_k}] - \tr[\rho {A_j}]\tr[\rho {A_k}] \in \mathbb{C},}
\end{array}}
\end{Equation}
{~\vspace{-24pt}\\}
\begin{Equation}                      {A.5}
{\begin{array}{*{20}{l}}
{\sigma _{{A_j}}^2}&\!\!{\equiv \sigma _{{A_j},{A_j}}^2 \equiv \langle {\psi _j}|{\psi _j}\rangle  \equiv {\mathop{\rm cov}} ({A_j},{A_j}) \equiv \tr[\rho {(\Delta {A_j})^2}]}\\
{}&\!\!{=\tr[\rho A_j^2] - {(\tr[\rho {A_j}])^2} \in \mathbb{R}\geq 0,}
\end{array}}
\end{Equation}
{~\vspace{-24pt}\\}
\begin{Equation}                      {A.6}
{\begin{array}{*{20}{l}}
{\sigma _{{A_j}}}&\!\!{\equiv \sqrt {\sigma _{{A_j}}^2}  \equiv \sqrt {\tr[\rho {{(\Delta {A_j})}^2}]}}\\
{}&\!\!{= \sqrt {\tr[\rho A_j^2] - {{(\tr[\rho {A_j}])}^2}}  \in \mathbb{R}\geq 0.}
\end{array}}
\end{Equation}
The \textit{generalized covariance}, \textit{generalized variance}, and \textit{generalized standard deviation} for mixed (not-necessarily normalized) states are
\begin{Equation}                      {A.7}
{\begin{array}{*{20}{l}}
{\widetilde{\sigma}_{{A_j},{A_k}}^2}&\!\!{\equiv \widetilde {{\mathop{\rm cov}} }({A_j},{A_k}) \equiv \tr[\rho {(\Delta {A_j})^\dag }(\Delta {A_k})]}\\
{}&\!\!{=\tr[\rho A_j^\dag {A_k}] - [2 - \tr(\rho )]\tr[\rho A_j^\dag ]\tr[\rho {A_k}] \in \mathbb{C},}
\end{array}}
\end{Equation}
{~\vspace{-24pt}\\}
\begin{Equation}                      {A.8}
{\begin{array}{*{20}{l}}
{\widetilde{\sigma}_{{A_j}}^2}&\!\!{\equiv \widetilde{\sigma}_{{A_j},{A_j}}^2 \equiv \widetilde {{\mathop{\rm cov}} }({A_j},{A_j}) \equiv \tr[\rho {(\Delta {A_j})^\dag }(\Delta {A_j})]}\\
{}&\!\!{= \tr[\rho A_j^\dag {A_j}] - [2 - \tr(\rho )]\tr[\rho A_j^\dag ]\tr[\rho {A_j}] \in \mathbb{R}\geq 0,}
\end{array}}
\end{Equation}
{~\vspace{-24pt}\\}
\begin{Equation}                      {A.9}
{\begin{array}{*{20}{l}}
{\widetilde{\sigma}_{{A_j}}}&\!\!{\equiv \sqrt {\widetilde{\sigma}_{{A_j}}^2}  \equiv \sqrt {\tr[\rho {{(\Delta {A_j})}^\dag }(\Delta {A_j})]}}\\
{}&\!\!{=\sqrt {\tr[\rho A_j^\dag {A_j}] - [2 - \tr(\rho )]\tr[\rho A_j^\dag ]\tr[\rho {A_j}]}  \in \mathbb{R}\geq 0.}
\end{array}}
\end{Equation}

If needed, similar results can be obtained for the regular vector versions by starting from the form for a single vector, as given above \Eq{7} as $\langle {A}\rangle  \!\equiv\! {\mathbf{a}^\dag }{A}\mathbf{a}=\tr(\mathbf{a}\mathbf{a}^{\dag}A)$, and defining the outer product of $\mathbf{a}$ as $\rho\equiv\mathbf{a}\mathbf{a}^{\dag}$ for a ``pure-vector density operator.'' Then by defining a ``mixed-vector density operator'' $\rho  \equiv \sum\nolimits_j {{p_j}{{\bf{a}}_{\{ j\} }}{\bf{a}}_{\{ j\} }^\dag } $ as some convex sum of outer products of pure decomposition vectors $\mathbf{a}_{\{j\}}$ with probabilities $p_j$ such that $\sum\nolimits_j {{p_j}}  = 1$, then the ``mixed-vector'' version of expectation values for regular complex vectors is
\begin{Equation}                      {A.10}
\begin{array}{*{20}{l}}
{\langle A\rangle}&\!\!{\equiv \tr(\rho A) = \tr(\sum\nolimits_j {{p_j}{{\bf{a}}_{\{ j\} }}{\bf{a}}_{\{ j\} }^\dag A} )}\\
{}&\!\!{= \sum\nolimits_j {{p_j}\tr({{\bf{a}}_{\{ j\} }}{\bf{a}}_{\{ j\} }^\dag A)}  = \sum\nolimits_j {{p_j}{\bf{a}}_{\{ j\} }^\dag A{{\bf{a}}_{\{ j\} }}},}\\
\end{array}
\end{Equation}
from which regular-vector analogues of all other quantities can be constructed for mixed-vector versions.\\

{\noindent}{\textit{Proof that All of These Results Hold for Mixed States}:}\\

First, it is well-known that the two-operator Schr{\"o}dinger relation from \Eq{1} is also true for mixed states, and is therefore also a demonstration that the CS inequality holds for convex sums of outer products of vectors.

Then, since the balanced and unbalanced multi-operator CS inequalities and uncertainty relations, in all their forms [including \Eq{2}, \Eq{5}, \Eq{11}, \Eq{12}, \Eq{14}, \Eq{20}, \Eq{23}, and \Eq{30}, and all corresponding quantities in the proofs and derivations in \Sec{III}], are based on products of the two-operator CS inequalities, or equivalently, of the two-operator Schr{\"o}dinger relation, then all of these results inherit the validity for mixed states as well (in particular because the inequalities all face the same way in such products).

For the results involving the multivariance, such as in \Eq{7}, \Eq{8}, \Eq{13}, all related quantities in the proofs and derivations in \Sec{III}, and in its symmetrized forms in the Conclusions in \Sec{V}, since they are merely an application of the two-operator CS inequality or the two-operator Schr{\"o}dinger relation to various different bipartitions of a given $M$-operator product of deviations, then all multivariance results are also valid for mixed states by the same principle.

Similarly, since the multi-operator squeezing topic of \Sec{IV} was presented as an application of the balanced multi-operator uncertainty relations, all of these results are also valid for mixed states by the same arguments above, including the brief summary given in \Sec{II.F}.

Therefore, rigorous proof that all results of this paper hold for mixed states all boils down to whether the original CS inequality holds for mixed states, or equivalently, to whether the two-operator Schr{\"o}dinger relation holds for mixed states.  Despite this being already well-known to be true, we will present a simple proof for it here, for completeness.

To do this efficiently, simply start with our generalized version of the Schr{\"o}dinger relation for pure states and not-necessarily Hermitian operators, written as
\begin{widetext}~\\
\vspace{-40pt}\\
\begin{Equation}                      {A.11}
\begin{array}{*{20}{rl}}
{{{\widetilde{\sigma}}_A}{{\widetilde{\sigma} }_B}}&\!\!{ \ge \left| {\langle \psi |{{(A - \langle A\rangle )}^\dag }(B - \langle B\rangle )|\psi \rangle } \right|}\\
{\sqrt {\tr[|\psi \rangle \langle \psi |{{(\Delta A)}^\dag }(\Delta A)]\tr[|\psi \rangle \langle \psi |{{(\Delta B)}^\dag }(\Delta B)]} }&\!\!{ \ge \left| {\tr[|\psi \rangle \langle \psi |{{(\Delta A )}^\dag }(\Delta B)]} \right|.}
\end{array}
\end{Equation}
Now relabel this pure state $|\psi \rangle$ as the $q$th pure state $|{\psi _{\{ q\} }}\rangle $ of some decomposition of a mixed state $\rho$, and multiply both sides by probability $p_q$ to obtain
\begin{Equation}                      {A.12}
{p_q}\sqrt {\langle {\psi _{\{ q\} }}|{{(\Delta A)}^\dag }(\Delta A)|{\psi _{\{ q\} }}\rangle \langle {\psi _{\{ q\} }}|{{(\Delta B)}^\dag }(\Delta B)]|{\psi _{\{ q\} }}\rangle }  \ge {p_q}\left| {\tr[|{\psi _{\{ q\} }}\rangle \langle {\psi _{\{ q\} }}|{{(\Delta A )}^\dag }(\Delta B )]} \right|,
\end{Equation}
and summing over $q$, this gives
\begin{Equation}                      {A.13}
\sum\nolimits_q {{p_q}\sqrt {\langle {\psi _{\{ q\} }}|{{(\Delta A)}^\dag }(\Delta A)|{\psi _{\{ q\} }}\rangle \langle {\psi _{\{ q\} }}|{{(\Delta B)}^\dag }(\Delta B)]|{\psi _{\{ q\} }}\rangle } }  \ge \sum\nolimits_q {{p_q}\left| {\tr[|{\psi _{\{ q\} }}\rangle \langle {\psi _{\{ q\} }}|{{(\Delta A  )}^\dag }(\Delta B  )]} \right|.} 
\end{Equation}
Now note that by the triangle inequality, for the quantity on the right in \Eq{A.13},
\begin{Equation}                      {A.14}
\begin{array}{*{20}{l}}
{\sum\nolimits_q {{p_q}\left| {\tr[|{\psi _{\{ q\} }}\rangle \langle {\psi _{\{ q\} }}|{{(\Delta A )}^\dag }(\Delta B )]} \right|} }&\!\!{ \ge \left| {\sum\nolimits_q {{p_q}\tr[|{\psi _{\{ q\} }}\rangle \langle {\psi _{\{ q\} }}|{{(\Delta A  )}^\dag }(\Delta B  )]} } \right|}\\
{\sum\nolimits_q {{p_q}\left| {\tr[|{\psi _{\{ q\} }}\rangle \langle {\psi _{\{ q\} }}|{{(\Delta A )}^\dag }(\Delta B )]} \right|} }&\!\!{ \ge \left| {\tr[\sum\nolimits_q {{p_q}|{\psi _{\{ q\} }}\rangle \langle {\psi _{\{ q\} }}|{{(\Delta A )}^\dag }(\Delta B )]} } \right|}\\
{\sum\nolimits_q {{p_q}\left| {\tr[|{\psi _{\{ q\} }}\rangle \langle {\psi _{\{ q\} }}|{{(\Delta A )}^\dag }(\Delta B )]} \right|} }&\!\!{ \ge \left| {\tr[\rho {{(\Delta A )}^\dag }(\Delta B )]} \right|,}
\end{array}
\end{Equation}
so putting \Eq{A.14} into \Eq{A.13} gives
\begin{Equation}                      {A.15}
\sum\nolimits_q {{p_q}\sqrt {\langle {\psi _{\{ q\} }}|{{(\Delta A)}^\dag }(\Delta A)|{\psi _{\{ q\} }}\rangle \langle {\psi _{\{ q\} }}|{{(\Delta B)}^\dag }(\Delta B)]|{\psi _{\{ q\} }}\rangle } }  \ge \left| {\tr[\rho {{(\Delta A )}^\dag }(\Delta B )]} \right|.
\end{Equation}
\end{widetext}
Here, since the whole quantity ${(\Delta A)^\dag }(\Delta A)$ is Hermitian (regardless of whether $\Delta A$ is or not), then the generalized variances such as $\langle {\psi _{\{ q\} }}|{(\Delta A)^\dag }(\Delta A)|{\psi _{\{ q\} }}\rangle $ are \textit{real} (because they are an expectation value of a Hermitian operator). Furthermore, if the singular value decomposition of $\Delta A$ is $\Delta A = U\Sigma {V^\dag }$ where $U$ and $V$ are unitary and $\Sigma$ is the diagonal matrix of its real, nonnegative singular values, then we see that the eigenvalues of ${(\Delta A)^\dag }(\Delta A)$ must be real and nonnegative, because its spectral decomposition is
\begin{Equation}                      {A.16}
\begin{array}{*{20}{l}}
{{(\Delta A)^\dag }(\Delta A)}&\!\!{= {(U\Sigma {V^\dag })^\dag }U\Sigma {V^\dag } = {(\Sigma {V^\dag })^\dag }{U^\dag }U\Sigma {V^\dag }}\\
{}&\!\!{= {V^\dag }^\dag {\Sigma ^\dag }\Sigma {V^\dag } = V\Sigma \Sigma {V^\dag } = V{\Sigma ^2}{V^\dag },}
\end{array}
\end{Equation}
where ${\Sigma ^2} \ge 0$. Thus, \smash{$\widetilde{\sigma}_A^2$} is an expectation value of a positive semidefinite Hermitian operator, so it is a real nonnegative number, with the same result holding for \smash{${\sigma}_A^2$}. Thus, letting ${a_q} \equiv \langle {\psi _{\{ q\} }}|{(\Delta A)^\dag }(\Delta A)|{\psi _{\{ q\} }}\rangle  \ge 0$ and ${b_q} \equiv \langle {\psi _{\{ q\} }}|{(\Delta B)^\dag }(\Delta B)]|{\psi _{\{ q\} }}\rangle  \ge 0$, \Eq{A.15} becomes
\begin{Equation}                      {A.17}
\sum\nolimits_q {{p_q}\sqrt {{a_q}{b_q}} }  \ge \left| {\tr[\rho {{(\Delta A )}^\dag }(\Delta B )]} \right|.
\end{Equation}

To show that the uncertainty product is always greater or equal to the left side of \Eq{A.17}, given nonequal integer index values $x \ne y$, for ${a_q},{b_q} \in \mathbb{Re} \ge 0$, it is true that
\begin{Equation}                      {A.18}
\begin{array}{*{20}{rl}}
{{{(\sqrt {{a_x}{b_y}}  - \sqrt {{a_y}{b_x}} )}^2}}&\!\!{ \ge 0}\\
{{a_x}{b_y} - 2\sqrt {{a_x}{b_y}{a_y}{b_x}}  + {a_y}{b_x}}&\!\!{ \ge 0}\\
{{a_x}{b_y} + {a_y}{b_x}}&\!\!{ \ge 2\sqrt {{a_x}{b_y}{a_y}{b_x}}. }
\end{array}
\end{Equation}
Then, multiply \Eq{A.18} by ${p_x}{p_y} \ge 0$ and rearrange as
\begin{Equation}                      {A.19}
\scalemath{0.94}{\begin{array}{*{20}{rl}}
{{p_x}{p_y}({a_x}{b_y} + {a_y}{b_x})}&\!\!{ \ge 2{p_x}{p_y}\sqrt {{a_x}{b_y}{a_y}{b_x}} }\\
{{p_x}{p_y}{a_x}{b_y} + {p_x}{p_y}{a_y}{b_x}}&\!\!{ \ge {p_x}{p_y}\sqrt {{a_x}{b_y}{a_y}{b_x}}  + {p_x}{p_y}\sqrt {{a_x}{b_y}{a_y}{b_x}} }\\
{{p_x}{p_y}{a_x}{b_y} + {p_y}{p_x}{a_y}{b_x}}&\!\!{ \ge {p_x}{p_y}\sqrt {{a_x}{b_x}{a_y}{b_y}}  + {p_y}{p_x}\sqrt {{a_y}{b_y}{a_x}{b_x}}. }
\end{array}}\!
\end{Equation}
Now sum over all such pairs of mismatched indices as
\begin{widetext}~
\begin{Equation}                      {A.20}
{\sum\nolimits_q {\sum\nolimits_r {(1 - {\delta _{q,r}}){p_q}{p_r}{a_q}{b_r}} }  \ge \sum\nolimits_q {\sum\nolimits_r {(1 - {\delta _{q,r}}){p_q}{p_r}\sqrt {{a_q}{b_q}{a_r}{b_r}} } }} .
\end{Equation}
Then noticing that the index-matched $q = r$ version of both sides of \Eq{A.20} without the $(1 - {\delta _{q,r}})$ switch is the same quantity \smash{$\sum\nolimits_q {p_q^2{a_q}{b_q}}$}, adding that to both sides and unifying both resulting sums gives
\begin{Equation}                      {A.21}
\begin{array}{*{20}{rl}}
{\displaystyle \sum\nolimits_q {p_q^2{a_q}{b_q}}  + \sum\nolimits_q {\sum\nolimits_r {(1 - {\delta _{q,r}}){p_q}{p_r}{a_q}{b_r}} } }&\!\!{\displaystyle \ge \sum\nolimits_q {p_q^2{a_q}{b_q}}  + \sum\nolimits_q {\sum\nolimits_r {(1 - {\delta _{q,r}}){p_q}{p_r}\sqrt {{a_q}{b_q}{a_r}{b_r}} } } }\\[10pt]
{\displaystyle \sum\nolimits_q {\sum\nolimits_r {{p_q}{p_r}{a_q}{b_r}} } }&\!\!{\displaystyle \ge \sum\nolimits_q {\sum\nolimits_r {{p_q}{p_r}\sqrt {{a_q}{b_q}{a_r}{b_r}} } }. }
\end{array}
\end{Equation}
\end{widetext}
This allows \Eq{A.21} to factor as
\begin{Equation}                      {A.22}
\begin{array}{*{20}{rl}}
{\left( {\sum\nolimits_q {{p_q}{a_q}} } \right)\left( {\sum\nolimits_{r\vphantom{q}} {{p_r}{b_r}} } \right)}&\!\!{ \ge \left( {\sum\nolimits_q {{p_q}\sqrt {{a_q}{b_q}} } } \right)\left( {\sum\nolimits_{r\vphantom{q}} {{p_r}\sqrt {{a_r}{b_r}} } } \right)}\\
{\left( {\sum\nolimits_q {{p_q}{a_q}} } \right)\left( {\sum\nolimits_{r\vphantom{q}} {{p_r}{b_r}} } \right)}&\!\!{ \ge {{\left( {\sum\nolimits_q {{p_q}\sqrt {{a_q}{b_q}} } } \right)}^2}}\\
{\sqrt {\sum\nolimits_q {{p_q}{a_q}} } \sqrt {\sum\nolimits_{r\vphantom{q}} {{p_r}{b_r}} } }&\!\!{ \ge \sum\nolimits_q {{p_q}\sqrt {{a_q}{b_q}} } .}
\end{array}\!\!
\end{Equation}
Then, noticing that
\begin{Equation}                      {A.23}
\begin{array}{*{20}{rl}}
{\sqrt {\sum\nolimits_q {{p_q}{a_q}} } }&\!\!{ = \sqrt {\sum\nolimits_q {{p_q}\langle {\psi _{\{ q\} }}|{{(\Delta A)}^\dag }(\Delta A)|{\psi _{\{ q\} }}\rangle } } }\\
{}&\!\!{ = \sqrt {\sum\nolimits_q {{p_q}\tr[|{\psi _{\{ q\} }}\rangle \langle {\psi _{\{ q\} }}|{{(\Delta A)}^\dag }(\Delta A)]} } }\\
{}&\!\!{ = \sqrt {\tr[\sum\nolimits_q {{p_q}|{\psi _{\{ q\} }}\rangle \langle {\psi _{\{ q\} }}|{{(\Delta A)}^\dag }(\Delta A)]} } }\\
{}&\!\!{ = \sqrt {\tr[\rho {{(\Delta A)}^\dag }(\Delta A)]}, }
\end{array}
\end{Equation}
which is the generalized standard deviation of $A$ for a system in mixed state $\rho$, and similarly for \rule{0pt}{11.5pt}\smash{$\sqrt {\rule{0pt}{8.5pt}\sum\nolimits_r {{p_r}{b_r}} }  = \sqrt {\rule{0pt}{9pt}\tr[\rho \smash{{{(\Delta B)}^\dag }(\Delta B)}]} $}, then \Eq{A.22} becomes
\begin{Equation}                      {A.24}
\sqrt {\tr[\rho {{(\Delta A)}^\dag }(\Delta A)]} \sqrt {\tr[\rho {{(\Delta B)}^\dag }(\Delta B)]}  \ge \sum\nolimits_q {{p_q}\sqrt {{a_q}{b_q}} }, 
\end{Equation}
and putting \Eq{A.24} into \Eq{A.17}, we get
\begin{Equation}                      {A.25}
\scalemath{0.94}{\sqrt {\tr[\rho {{(\Delta A)}^\dag }(\Delta A)]} \sqrt {\tr[\rho {{(\Delta B)}^\dag }(\Delta B)]}  \ge\! \left| {\tr[\rho {{(\Delta A )}^\dag }(\Delta B )]} \right|},
\end{Equation}
which is the result we set out to prove for mixed states, which is the generalization of the Schr{\"o}dinger relation for general operators and generally mixed states, more compactly written as
\begin{Equation}                      {A.26}
{\widetilde{\sigma}_A}{\widetilde{\sigma}_B} \ge |\widetilde{\sigma}_{A,B}^2|.
\end{Equation}
Note that \Eq{A.25} and \Eq{A.26} simplify to the mixed-state Schr{\"o}dinger relation when $A$ and $B$ are Hermitian.  Furthermore, this proves the analogous results for special case versions such as the Robertson relation.

Then, since all of our other results are product compositions of \Eq{A.25} with the inequalities all facing the same way, all of our results in this paper still hold for mixed states by rewriting all expectation values as traces involving an outer-product of a pure state which can then be directly replaced by a mixed-state density operator.

Tighter relationships can of course be obtained (such as with convex-roof extensions, etc.), but as we have explained, directly computing the uncertainty product is already the tightest relation, so the simple method of upgrading a pure $\rho$ to a mixed $\rho$ is beneficial precisely because of its simplicity.
\end{appendix}
%
\end{document}